\documentclass[twocolumn, floatfix]{aastex631}

\usepackage{ulem}
\usepackage{float}
\usepackage{gensymb}
\usepackage{multirow}
\usepackage{longtable}
\usepackage{threeparttable} 
\usepackage{hyperref}
\usepackage{amsmath} 

\usepackage{breqn}
\usepackage[flushmargin]{footmisc}

\usepackage{tikz}
\usetikzlibrary{calc,tikzmark} 
\newlength{\imageheight}
\newcommand{\FontLabel}{\sffamily\bfseries\large}

\newcommand{\InsertLabels}[2]{%
  \begin{tikzpicture}[overlay,remember picture]
        \node[font=\FontLabel] at ( $ (pic cs:a) +(#1,\imageheight-#2) $ ){a};%
        \node[font=\FontLabel] at ( $ (pic cs:b) +(#1,\imageheight-#2) $ ){b};%
        \node[font=\FontLabel] at ( $ (pic cs:c) +(#1,\imageheight-#2-0.5cm) $ ){c};%
        \node[font=\FontLabel] at ( $ (pic cs:d) +(#1,\imageheight-#2-0.5cm) $ ){d};%
    \end{tikzpicture}   
}

\shorttitle{MAGIC metallicities for the Sculptor dwarf spheroidal galaxy}
\shortauthors{Barbosa et al.}

\graphicspath{{./}{figures/}}

\begin{document}

\title{The DECam MAGIC Survey: A Wide-field Photometric Metallicity Study of the Sculptor Dwarf Spheroidal Galaxy}

\correspondingauthor{F. O. Barbosa}
\email{fabriciaob@usp.br}

\author[0000-0002-8262-2246]{F.~O.~Barbosa}
\affiliation{Universidade de S\~ao Paulo, Instituto de Astronomia, Geof\'isica e Ci\^encias Atmosf\'ericas, Departamento de Astronomia, SP 05508-090, S\~ao Paulo, Brasil}

\author[0000-0002-7155-679X]{A.~Chiti}
\affiliation{Department of Astronomy \& Astrophysics, University of Chicago, 5640 S. Ellis Avenue, Chicago, IL 60637, USA}
\affiliation{Kavli Institute for Cosmological Physics, University of Chicago, 5640 S. Ellis Avenue, Chicago, IL 60637, USA}

\author[0000-0002-9269-8287]{G.~Limberg}
\affiliation{Kavli Institute for Cosmological Physics, University of Chicago, 5640 S. Ellis Avenue, Chicago, IL 60637, USA}

\author[0000-0002-6021-8760]{A.~B.~Pace}
\affiliation{Department of Astronomy, University of Virginia, 530 McCormick Road, Charlottesville, VA 22904, USA}

\author[0000-0003-1697-7062]{W.~Cerny}
\affiliation{Department of Astronomy, Yale University, New Haven, CT 06520, USA}

\author[0000-0001-7479-5756]{S.~Rossi}
\affiliation{Universidade de S\~ao Paulo, Instituto de Astronomia, Geof\'isica e Ci\^encias Atmosf\'ericas, Departamento de Astronomia, SP 05508-090, S\~ao Paulo, Brasil}

\author[0000-0002-3936-9628]{J.~L.~Carlin}
\affiliation{Rubin Observatory/AURA, 950 North Cherry Avenue, Tucson, AZ, 85719, USA}

\author[0000-0003-1479-3059]{G.~S.~Stringfellow}
\affiliation{Center for Astrophysics and Space Astronomy, University of Colorado Boulder, Boulder, CO 80309, USA}

\author[0000-0003-4479-1265]{V.~M.~Placco}
\affiliation{NSF NOIRLab, 950 N. Cherry Ave., Tucson, AZ 85719, USA}

\author[0000-0001-9649-8103]{K.~Atzberger}
\affiliation{Department of Astronomy, University of Virginia, 530 McCormick Road, Charlottesville, VA 22904, USA}

\author[0000-0002-3690-105X]{J.~A.~Carballo-Bello}
\affiliation{Instituto de Alta Investigaci\'on, Universidad de Tarapac\'a, Casilla 7D, Arica, Chile}

\author[0000-0001-5143-1255]{A.~Chaturvedi}
\affiliation{Department of Physics, University of Surrey, Guildford GU2 7XH, UK}

\author[0000-0003-1680-1884]{Y.~Choi}
\affiliation{NSF NOIRLab, 950 N. Cherry Ave., Tucson, AZ 85719, USA}

\author[0000-0002-1763-4128]{D.~Crnojevi\'c}
\affiliation{Department of Physics \& Astronomy, University of Tampa, 401 West Kennedy Boulevard, Tampa, FL 33606, USA}

\author[0000-0001-8251-933X]{A.~Drlica-Wagner}
\affiliation{Fermi National Accelerator Laboratory, P.O.\ Box 500, Batavia, IL 60510, USA}
\affiliation{Kavli Institute for Cosmological Physics, University of Chicago, 5640 S. Ellis Avenue, Chicago, IL 60637, USA}
\affiliation{Department of Astronomy \& Astrophysics, University of Chicago, 5640 S. Ellis Avenue, Chicago, IL 60637, USA}
\affiliation{NSF-Simons AI Institute for the Sky (SkAI),172 E. Chestnut St., Chicago, IL 60611, USA}

\author[0000-0002-4863-8842]{A.~P.~Ji}
\affiliation{Department of Astronomy \& Astrophysics, University of Chicago, 5640 S. Ellis Avenue, Chicago, IL 60637, USA}
\affiliation{Kavli Institute for Cosmological Physics, University of Chicago, 5640 S. Ellis Avenue, Chicago, IL 60637, USA}

\author[0000-0002-3204-1742]{N.~Kallivayalil}
\affiliation{Department of Astronomy, University of Virginia, 530 McCormick Road, Charlottesville, VA 22904, USA}

\author[0000-0002-9144-7726]{C.~E.~Mart\'inez-V\'azquez}
\affiliation{NSF NOIRLab, 670 N. A'ohoku Place, Hilo, Hawai'i, 96720, USA}

\author[0000-0003-0105-9576]{G.~E.~Medina}
\affiliation{David A. Dunlap Department of Astronomy \& Astrophysics, University of Toronto, 50 St George Street, Toronto ON M5S 3H4, Canada}
\affiliation{Dunlap Institute for Astronomy \& Astrophysics, University of Toronto, 50 St George Street, Toronto, ON M5S 3H4, Canada}

\author[0000-0002-8282-469X]{N.~E.~D.~No\"el}
\affiliation{Department of Physics, University of Surrey, Guildford GU2 7XH, UK}

\author[0000-0001-5805-5766]{A.~H.~Riley}
\affiliation{Institute for Computational Cosmology, Department of Physics, Durham University, South Road, Durham DH1 3LE, UK}

\author[0000-0003-4102-380X]{D.~J.~Sand}
\affiliation{Department of Astronomy/Steward Observatory, 933 North Cherry Avenue, Room N204, Tucson, AZ 85721-0065, USA}

\author[0000-0003-4341-6172]{A.~K.~Vivas}
\affiliation{Cerro Tololo Inter-American Observatory/NSF NOIRLab, Casilla 603, La Serena, Chile}

\author[0000-0003-4383-2969]{C.~R.~Bom}
\affiliation{Centro Brasileiro de Pesquisas F\'isicas, Rua Dr. Xavier Sigaud 150, 22290-180 Rio de Janeiro, RJ, Brazil}

\author[0000-0001-6957-1627]{P.~S.~Ferguson}
\affiliation{DiRAC Institute, Department of Astronomy, University of Washington, 3910 15th Ave NE, Seattle, WA, 98195, USA}

\author[0000-0001-9649-4815]{B.~Mutlu-Pakdil}
\affiliation{Department of Physics and Astronomy, Dartmouth College, Hanover, NH 03755, USA}

n\author[0000-0001-9438-5228]{M.~Navabi}
\affiliation{Department of Physics, University of Surrey, Guildford GU2 7XH, UK}

\author[0000-0002-1594-1466]{J.~D.~Sakowska}
\affiliation{Department of Physics, University of Surrey, Guildford GU2 7XH, UK}
\affiliation{Instituto de Astrofísica de Andalucía, CSIC, Glorieta de la Astronom\'\i a,  E-18080 Granada, Spain}

\author[0000-0001-6455-9135]{A.~Zenteno}
\affiliation{Cerro Tololo Inter-American Observatory/NSF NOIRLab, Casilla 603, La Serena, Chile}

\collaboration{30}{(MAGIC \& DELVE Collaborations)}

\begin{abstract}

The metallicity distribution function and internal chemical variations of a galaxy are fundamental to understand its formation and assembly history. In this work, we analyze photometric metallicities for 3883 stars over seven half-light radii ($\rm r_h$) in the Sculptor dwarf spheroidal (Scl dSph) galaxy, using new narrow-band imaging data from the Mapping the Ancient Galaxy in CaHK (MAGIC) survey conducted with the Dark Energy Camera (DECam) at the 4-m Blanco Telescope. This work demonstrates the scientific potential of MAGIC using the Scl dSph galaxy, one of the most well-studied satellites of the Milky Way. 
Our sample ranges from $\rm [Fe/H] \approx - 4.0$ to $\rm [Fe/H] \approx - 0.6$, includes six new extremely metal-poor candidates ($\rm [Fe/H] \leq -3.0$), and is almost three times larger than the largest spectroscopic metallicity dataset in the Scl dSph. Our spatially unbiased sample of metallicities provides a more accurate representation of the metallicity distribution function, revealing a more metal-rich peak than observed in the most recent spectroscopic sample. 
It also reveals a break in the metallicity gradient, with a strong change in the slope: from $-3.26 \pm 0.18 \rm \ dex \, deg^{-1}$ for stars inside $\sim 1\ \rm r_h$ to $-0.55 \pm 0.26 \rm \ dex \, deg^{-1}$ for the outer part of the Scl dSph. Our study demonstrates that combining photometric metallicity analysis with the wide field of view of DECam offers an efficient and unbiased approach for studying the stellar populations of dwarf galaxies in the Local Group.

\end{abstract}


\defcitealias{Tolstoy2023}{T23}
\defcitealias{delosReyes22}{R22}

\section{Introduction} \label{sec:intro}
\setcounter{footnote}{23}

According to the $\Lambda$ Cold Dark Matter paradigm, massive galaxies are formed by the aggregation of smaller systems (\citealt{Searle1978}, \citealt{Faber1979}, \citealt{White1991}, \citealt{Kauffmann1993}, \citealt{Johnston1998}, \citealt{Springel2006}). There is plenty of evidence, from kinematic and chemical properties, to suggest that the halo of the Milky Way is composed of a diverse range of disrupted dwarf galaxies (see reviews by \citealt{Helmi2008_rev}, \citealt{Belokurov2013_rev}, \citealt{Helmi2020_rev}, and \citealt{Deason2024_rev}). 
These disrupted systems were once bound to the Milky Way, similar to the satellites observed today in the Local Group. Therefore, the study of present-day surviving satellites can be used to study the formation and evolution of building blocks that form galactic halos, and also test galaxy formation processes in the low-mass regime. 

To understand the formation history and evolution of dwarf galaxies, the investigation of the chemical composition of individual member stars is desirable. Particularly, stellar metallicity distribution functions (MDFs) are useful to shed light on the mechanisms that shape the chemical evolution of these galaxies (\citealt{lai2011}, \citealt{Kirby2011, Kirby2015}). To ensure an accurate inference of these properties through fitting analytic chemical evolution models (e.g., \citealt{Weinberg2017}, \citealt{Sandford2024}) or simply in describing the metallicity distribution of a galaxy, it is crucial to analyze MDFs that are well-sampled ($\gtrsim$1000 stars; \citealt{Kordopatis2016}, \citealt{Pace2020}) and spatially unbiased. 

However, acquiring spectra for a large number of stars in dwarf galaxies is observationally expensive. Measurements of iron lines require medium/high-resolution spectroscopy ($R > 10,000$), which means a considerable amount of time (up to a few hours per object) observing even the brightest red giant branch stars (RGB) in systems as distant as the Milky Way's classical satellites (beyond 70 kpc; \citealt{McConnachie2012}, \citealt{Pace2024}). The first spectroscopic studies of Milky Way satellites observed less than a dozen stars (e.g., \citealt{Boni2000_T09}, \citealt{She2001_T09}). 
Following works, especially those with multi-object observations, increased the number of high-resolution spectra obtained to tens (e.g., \citealt{Monaco2005_T09}, \citealt{Koch2008_T09}, \citealt{Battaglia2008_VLT}, \citealt{Lardo2016}). 
Still, the current samples are generally limited to a few hundreds of stars for not even a handful of dwarf spheroidal galaxies (e.g., Sextans, Carina, Sculptor, Draco; \citealt{Kirby2009}, \citealt{Walker2009, Walker2023}).

To overcome this limitation, we can use photometric metallicities derived from narrow-band filters centered around metallicity-sensitive features of stellar spectra. Although this technique is less precise than spectroscopic metallicities, it has proved to be a valuable tool for estimating metallicities for larger samples of stars in dwarf galaxies with a higher efficiency (\citealt{Longeard2018, Longeard2020}, \citealt{Chiti2020}, \citealt{Vitali2022}, \citealt{Fu2022, Fu2023, Fu2024_Tuc, Fu2024_M31}, \citealt{Pan2025}). 
The Ca II H and K lines are the strongest metal lines in a stellar spectrum \citep{Beers1999}. Their strength can thus be measured photometrically with a narrow filter around $\sim 395$ nm (hereafter referred to as CaHK filter). At fixed stellar parameters (e.g., effective temperature, surface gravity), a smaller observed flux on the CaHK filter indicates higher element abundances, as more metal-rich stars ought to have stronger Ca II absorption lines.

The effectiveness of CaHK observations has been demonstrated in several previous works. 
Spectroscopic programs, such as the HK (\citealt{Beers1985}, \citealt{Beers1992}) and the Hamburg/ESO (\citealt{Frebel2006}, \citealt{Christlieb2008}) surveys, used the Ca II K line to identify metal-poor candidates (see \citealt{Limberg2021}). 
It has also been used in photometric studies, such as \cite{AnthonyT1991}, that introduced the Ca filter into the \textit{uvby} system to improve the loss of sensitivity below $\rm[Fe/H] = -2$ in previous metallicity indices. 
More recently, the SkyMapper Southern Sky survey \citep{skymapper} and the Pristine survey \citep{Starkenburg2017_pristine} have achieved significant results in the search for metal-poor stars in the Milky Way using metallicity-sensitive filters. The former introduced the \textit{v} filter, which encompasses the Ca II lines, while the latter employs a narrowband filter centered on the Ca II K line. These advances enabled the identification of a few hundred extremely metal-poor stars and have contributed to the characterization of metal-poor stellar systems, such as ultra-faint dwarf galaxies (e.g., \citealt{Keller2014}, \citealt{Youakim2017}, \citealt{Aguado2019}, \citealt{Costa2019}, \citealt{Chiti2020}, \citealt{Yong2021}, \citealt{Longeard2022}). 
Moreover, the potential of photometric metallicities was explored to estimate metallicity gradients for Milky Way satellites \citep{Han2020} using Subaru/Suprime-Cam \citep{suprimecam}, and for two M31 satellites with the Hubble Space Telescope \citep{Fu2024_M31}, which could not be done with ground-based spectroscopy. Ongoing narrow-band surveys, such as the Southern Photometric Local Universe Survey \citep[S-PLUS;][]{splus}, Javalambre Photometric Local Universe Survey \citep[J-PLUS;][]{jplus}, and Javalambre-Physics of the Accelerating Universe Astrophysical Survey \citep[J-PAS;][]{jpas} are also poised to obtain photometric metallicities in dwarf galaxies for samples of thousands of stars.

The present work explores data from a new photometric CaHK survey, the Mapping the Ancient Galaxy in CaHK (MAGIC; PI: Anirudh Chiti), conducted with the Dark Energy Camera (DECam; \citealt{Flaugher2015}), on the Victor Blanco 4-m telescope at Cerro Tololo Inter-American Observatory, in Chile, as part of the DECam Local Volume Exploration Survey (DELVE; \citealt{Drlica2021}). In particular, we analyze data for the Sculptor dwarf spheroidal galaxy (Scl dSph), one of the most extensively studied dwarf galaxies in the Local Group, providing numerous spectroscopic studies for comparison (e.g., \citealt{Kirby2009}, \citealt{Simon2015}, \citealt{Chiti2018}, \citealt{Hill2019}, \citealt{delosReyes22}, \citealt{Walker2023}, \citealt{Tang2023}, \citealt{Tolstoy2023}). Its location is especially suitable for spectroscopic and photometric studies. It is found in the direction of Galactic pole ($b \sim - 83\degree$), less susceptible to the Galactic extinction, and is close to the Milky Way in comparison to other classical dwarf satellites (heliocentric distance of $83.9 \pm 1.5$ kpc; \citealt{Martinez2015}, \citealt{Pace2022}). 

In the first detailed studied of Scl dSph's star formation history, \cite{deBoer2012_mp} suggested an extended star formation period lasting approximately 6--7 Gyr, a range supported by later studies (\citealt{Martinez2016_grad}, \citealt{Savino2018}). More recent works, however, report a shorter period of star formation, between 0.9--2.2 Gyr (\citealt{Weisz2014}, \citealt{Vincenzo2016}, \citealt{Bettinelli2019}, \citealt{delosReyes22}). 
Despite the proposed brief period of star formation, two distinct chemo-dynamical populations characterize it. The central region is dominated by younger, more metal-rich stars with lower velocity dispersion, whereas the more metal-poor population extends from the inner to the outer regions and exhibits higher velocity dispersion (\citealt{Tolstoy2004}, \citealt{Arroyo2024}). Recently, a third, even lower-metallicity, population was potentially identified with both chemistry and kinematics \citep{Arroyo2024}. These updates highlight the importance of systematic, large chemodynamic studies of the Scl dSph, and dSphs more broadly, to understand the processes that shape galaxy formation at the low mass end. 

The aim of this paper is to analyze chemical properties of the Scl dSph with the new MAGIC survey using the DECam CaHK filter, while also testing the validity of wide-field photometric metallicities.
Considering all the properties of the Scl dSph that earned it the status of a ``textbook dSph galaxy" \citep{Hill2019}, we choose it as the test case to validate our photometric metallicities. 
Until recently, only 12 members with $\rm[Fe/H] < -2.5$ were known \citep{Sestito2023}. This number has increased to 74 with the latest analysis of Very Large Telescope (VLT) data by \cite{Tolstoy2023}. Our sample doubles the number of members of this previous work, with about 20 of them being extremely metal-poor (EMP; $\rm[Fe/H] < -3$).
We show that the photometric metallicities are reliable, potentially reaching [Fe/H] as low as $-4.0$ dex, and they allow us to re-derive the MDF and the metallicity gradient of the Scl dSph. 

This work demonstrates the power of the MAGIC survey for studies of the Milky Way's stellar halo \citep{Placco2025} and dwarf galaxies in the Local Group. Our catalog for the Scl dSph provides a spatially unbiased metallicity sample and is more complete than spectroscopic samples currently available. This sample enables more accurate studies of the structural characteristics of the Scl dSph, and a better definition of the MDF and the metallicity behavior into its outskirts.

This work is organized as follows. We describe briefly the catalogues used in Section \ref{sec:data}. The method implemented to derive the metallicities and the member selection are presented in Section \ref{sec:meth}. In Section \ref{sec:comp}, we evaluate the performance of the photometric metallicities by comparing with previous spectroscopic works. The implications of the new sample for the characterization of the Scl dSph are discussed in Section \ref{sec:res}. Finally, we summarize our results in Section \ref{sec:conc}.

\section{Data}
\label{sec:data}

Our photometric dataset combines the DECam broad-band data from DELVE and the new narrow-band CaHK observations from MAGIC. We describe each of these components below.

\subsection{The MAGIC Survey}
\label{magic}

The MAGIC survey (Chiti et al. in prep.) provides narrow-band photometry using a CaHK filter on DECam. 
The new filter, designated as N395, has a central wavelength of 395.2 nm, a full width at half maximum of 10 nm, and a ``top hat'' transmission profile. It was designed to encompass the Ca II H and K lines at 396.85 nm and 393.37 nm, respectively, replicating similar filters used for previous narrow-band photometric surveys (e.g., Pristine survey; \citealt{Starkenburg2017_pristine}). The measured flux through this filter can be combined with DELVE broad-band observations to infer the metallicities for stars over the wide field observed by the DECam. MAGIC aims to cover at least $\sim$5,500 deg$^2$ of the southern sky, including a high number of known Milky Way satellites, being a powerful tool for investigating their MDFs and the Milky Way itself. At this stage, MAGIC has covered $\sim$1,500 deg$^2$ and data collection is ongoing.

MAGIC will provide an extensive and homogeneous sample of metallicities, potentially reaching values as low as $\rm [Fe/H] = -4.0$. It will provide deeper CaHK imaging than most narrow-band CaHK surveys (e.g., the aforementioned Pristine and S/J-PLUS surveys) due to longer exposure times (e.g., 720\,s for MAGIC versus 100\,s for normal Pristine pointings; \citealt{Starkenburg2017_pristine}) and larger aperture ($\sim$4\,m for DECam versus $\sim$1\,m for S-PLUS; \citealt{splus}), allowing the identification of numerous metal-poor stars in the outer Milky Way halo, stellar streams and substructures in dwarf galaxies. 
The initial release of the MAGIC catalog is planned for 2026. It will include fields spanning part of the Milky Way halo, in addition to observations of globular clusters, other dSph and ultra-faint dwarf galaxies, and a portion of the outskirts of the Magellanic Clouds. 
We note that the data analyzed in this paper were obtained from the earliest runs of the MAGIC program (NOIRLab Prop. ID 2023B-646244; P.I. Anirudh Chiti). The field of view\footnote{A single DECam field.} covers about 7 half-light radii ($\rm r_h$) around the center of the Scl dSph galaxy, and was obtained with three exposures of 12 minutes, dithered slightly to cover chip gaps. 

The details regarding data reduction for the MAGIC survey will be presented in an upcoming early science summary paper (Chiti et al. in prep.); we summarize the reduction and photometry specific to the Sculptor dSph data here.
The Sculptor N395 images are available on the NOIRLab data archive\footnote{https://astroarchive.noirlab.edu}, with post-reduction by the DECam Community Pipeline \citep[CP;][]{vg+14}.
We retrieved the three CP-processed N395 exposures of Sculptor dSph, along with their associated CP-generated weight maps and data quality masks.
We generated source catalogs from each exposures using \texttt{Source Extractor} and \texttt{PSFEx} \citep{ba+96, b+11}, with configuration files for this software from the DES Data Management system pipeline (\citealt{Desai2012}, \citealt{Mohr2012}, \citealt{mgm+18}, \citealt{desdr2}).

The photometric zero-point offset was calculated by finding the median offset between instrumental N395 magnitudes and synthetic CaHK magnitudes obtained from flux-calibrated Gaia XP data \citep{gaiadr3,GaiaXP}, as has been done with previous narrow-band CaHK calibration (e.g., \citealt{Martin2024}, \citealt{Perottoni2024}, \citealt{Xiao2024}). 
We found 120 -- 140 stars in each Sculptor N395 exposure in common between the DECam data and the catalog of Gaia XP-based synthetic CaHK magnitudes with uncertainty $< 0.05$\,mag, which we utilized to perform this correction.
After zero-point correction, the exposure-level source catalogs were combined and repeated N395 magnitude measurements were combined via a weighted mean.

We propagate the standard flags\footnote{https://sextractor.readthedocs.io/en/latest/Flagging.html} from \texttt{Source Extractor} and \texttt{PSFEx} to our final catalog, into the \texttt{feh\_phot\_flag} parameter.
All sources with $\texttt{feh\_phot\_flag} \geq 4$ are excluded from this analysis, as this indicates that at least one pixel in the source was saturated. 
Sources with 
$\texttt{feh\_phot\_flag} = \{2, 3\}$ indicate neighboring sources that required de-blending; these objects are retained in our catalog, and we hereafter refer to these as having the \texttt{feh\_phot\_flag} indicator for being de-blended.
We discuss the metallicity performance for de-blended sources in Section~\ref{spec}, and find no systematic impact relative to sources that were not de-blended when comparing to literature spectroscopic metallicities.

\subsection{DELVE}
\label{delve}

DELVE (NOIRLab Prop. ID 2019A-0305; PI: Alex Drlica-Wagner), currently in its second data release (DR2\footnote{\url{https://datalab.noirlab.edu/delve/}}), is a survey program that aggregates DECam data imaged by different past programs, in addition to complementing coverage with new observations \citep{Drlica2021, Drlica2022}. The DELVE DR2 catalogue is assembled from data collected by DELVE itself \citep{Drlica2021}, Dark Energy Survey (DES; \citealt{des}) and DECam Legacy Surveys (DECaLS; \citealt{Dey2019_desi}), besides other observing programs using the DECam, providing $griz$ photometry for $\sim$ 17,000 deg$^2$ (around 618 million objects) with $5\sigma$ point source depths of $g = 24.3$, $r = 23.9$, $i = 23.5$, and $z = 22.8$ mag, and a photometric accuracy of $\lesssim 20$ mmag. The mission focus is on characterizing Milky Way satellites and other stellar systems in the Local Group, as well as the discovery of new dwarf galaxies. Over the course of four years of work, numerous ultra-faint dwarf galaxies have been discovered using DELVE data (\citealt{Mau2020_ufd}, \citealt{Cerny2021_ufd, Cerny2023_ufd, Cerny2024_ufd}, \citealt{Tan2024_ufd}).

\section{Methodology}
\label{sec:meth}

The method implemented to compute photometric metallicities with the CaHK filter is similar to what was presented in \cite{Chiti2020, Chiti2021}, which we describe in this section. The notable difference is that these studies used the SkyMapper \textit{v} filter \citep{skymapper} to derive metallicities, whereas here we adapt the technique to use the narrow-band DECam CaHK filter.

\subsection{Photometric metallicities}
\label{feh}

As seen in previous studies that made use of metallicity-sensitive filters covering the CaHK region (e.g., \citealt{Costa2019}, \citealt{Chiti2020}, \citealt{Starkenburg2017_pristine}, \citealt{Huang2022}, \citealt{Placco2022}, \citealt{Almeida2023}), color-color plots combining the CaHK, $g$ and $i$ filters can be used to distinguish different metallicities for a given surface gravity value ($\log{\rm g}$). To derive metallicities in this work, we adopt the same approach as in \cite{Chiti2020, Chiti2021}, by forward-modelling fluxes through the CaHK, \textit{g}, and \textit{i} filters using a grid of flux-calibrated synthetic spectra spanning a range of stellar parameters. These forward-modeled fluxes are used to populate the aforementioned color-color space to map the color values to metallicities. Specifically, the grid of synthetic spectra (referred to as the TSLTE grid -- TurboSpectrum with Local Thermodynamic Equilibrium) is generated using the Turbospectrum code (\citealt{Alvarez1998_turbospec}, \citealt{Plez2012_turbospec}), MARCS model atmospheres (Model Atmospheres with a Radiative and Convective Scheme; \citealt{Gustafsson2008_marcsmodel}), and the line lists provided in VALD (Vienna Atomic Line Database; \citealt{Piskunov1995}, \citealt{Ryabchikova2015}).

\begin{figure}[t]
\begin{center}
\includegraphics[width=\columnwidth]{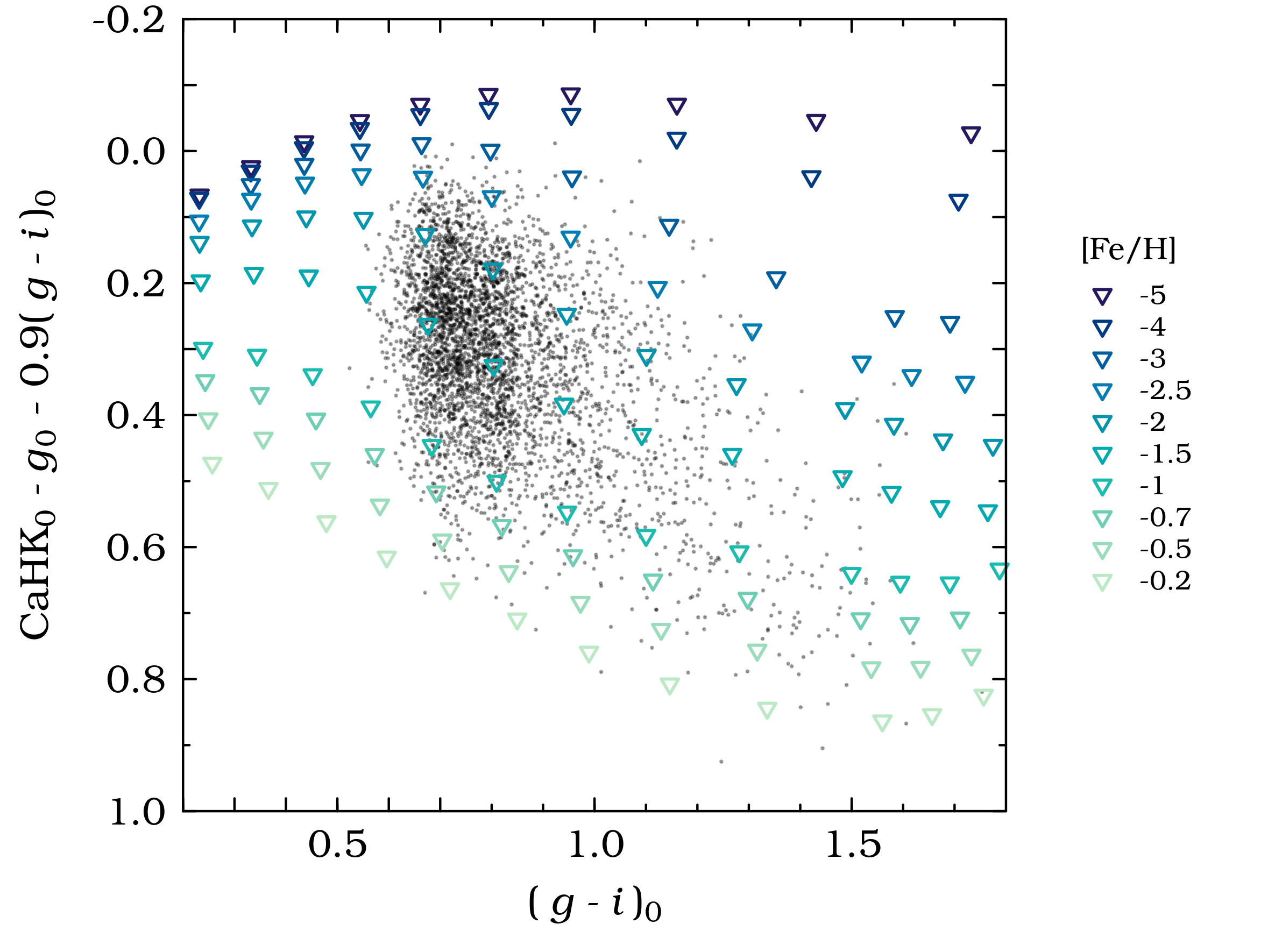}
\caption{A color-color plot of the MAGIC data on the Scl dSph galaxy (black data points), in which the y-axis color term includes the CaHK magnitude, and the x-axis is $(g-i)_0$ color. Stars at different metallicities occupy different regions of this color-color space. To indicate this, we overplot as upside-down triangles synthetic photometry from the TSLTE grid (described in Section \ref{feh}), fixing $\rm log \ g = 2.0$. As seen by the color-scale, stars at different metallicities separate into well-behaved trend-lines.}
\label{fig:models}
\end{center}
\end{figure}

The TSLTE grid consists of 13 values of $\log{\rm g}$ equally spaced between $\log{\rm g} = -0.5$ and $\log{\rm g} = +5.5$, and metallicities from $\rm [Fe/H] = -5$ to $\rm [Fe/H] = +1$. The color-color space generated by these models in the space ${\rm CaHK_0} - g_0 + 0.9\cdot (g - i)_0$ versus $(g - i)_0$, as shown on Figure \ref{fig:models}, are interpolated using \texttt{scipy.interpolate.grid-data} \citep{2020SciPy}. The subscript signifies magnitudes corrected for reddening with the \cite{Schlegel1998} dust map. 
For CaHK magnitudes, we adopted the extinction coefficient from \cite{Starkenburg2017_pristine}: $A_{\rm CaHK}/E(B-V) = 3.924$. The detailed process of the metallicity determination will be described in Chiti et al. (in prep.).

For each value of $\log{\rm g}$, the location of the models in the color-color plot shifts slightly. Therefore, it is necessary to choose an initial $\log{\rm g}$ value for the metallicity calculation. For the purpose of this work, where the distance to the stars is well-known as the distance of the Scl dSph galaxy, the initial guess of $\log{\rm g}$ is derived from basic assumptions. We used the color--effective temperature transformation for giants with color $(G - G_{\rm RP})_0$ provided by \cite{Mucciarelli2021_color-teff} for Gaia DR3 photometry (\citealt{Gaia}, \citealt{gaiadr3_phot}, \citealt{gaiadr3}), neglecting the dependence on the metallicity, since it does not have a significant impact in the final estimates\footnote{Adopting an initial value of $\rm[Fe/H] = -1.8$ would decrease the final metallicities by $\sim 0.02$ dex.}. The luminosities are estimated using the absolute magnitude of the Sun in DES \textit{g}-band provided by \cite[][$M_g = 5.05$]{Willmer2018_sunmag} and the bolometric corrections derived with \texttt{isochrones} Python package \citep{isoc_pck_bc}, adopting [Fe/H] = -1.8 and $\log \ \rm g = 2.0$. Then, the stellar radii are calculated with Stefan-Boltzmann relation assuming a mass of $0.78 \ M_{\odot}$ for the RGB stars (\citealt{Howell2024}) to obtain the $\log{\rm g}$. 

The statistical uncertainty in the metallicity is derived by propagating the uncertainty in the CaHK, \textit{g}, and \textit{i} filters, and re-deriving the metallicity. The systematic uncertainty floor in the metallicity from this method is assumed to be 0.16 dex, following \cite{Chiti2020}, who implemented the same technique and compared results to globular cluster metallicities in their study. 
This metallicity uncertainty floor also produces errors that are reasonably consistent with literature spectroscopic studies (see Section~\ref{spec}), indicating this is an appropriate value.

\subsection{Scl dSph members}
\label{mem}

Our selection of members in the Scl dSph galaxy follows the catalog presented in \cite{Pace2022}, which uses the proper motion, spatial distribution, a color-magnitude diagram selection using DECam photometry and Dartmouth isochrones \citep{Dotter2008}, and an additional color–magnitude box selection with Gaia photometry to assign a membership probability to each star. 
We adopted the membership probability estimated for their complete sample with a uniform background model. This sample was obtained selecting stars based on Gaia EDR3 \citep{gaiaedr3} data: $G < 21$, renormalized unit-weight error (RUWE) $< 1.4$, parallax $\varpi - 3.5\sigma_{\varpi} < 0$, and $v_{tan} - 3.5\sigma_{v_{tan}} < v_{esc}$, with the tangential ($v_{tan}$) and escape ($v_{esc}$) velocities calculated as described in \cite{Pace2022}. 
We chose to limit the sample to stars with a membership probability greater than $0.80$. We note that the selection of member stars is not particularly sensitive to the membership probability threshold that we apply, as the majority (6057 from 8768 stars in \citeauthor{Pace2022}'s sample) have membership probabilities above $0.99$.

\begin{figure*}[pt!]
\centering
  \begin{minipage}[b]{\columnwidth}
    \centering
    \includegraphics[width=\columnwidth]{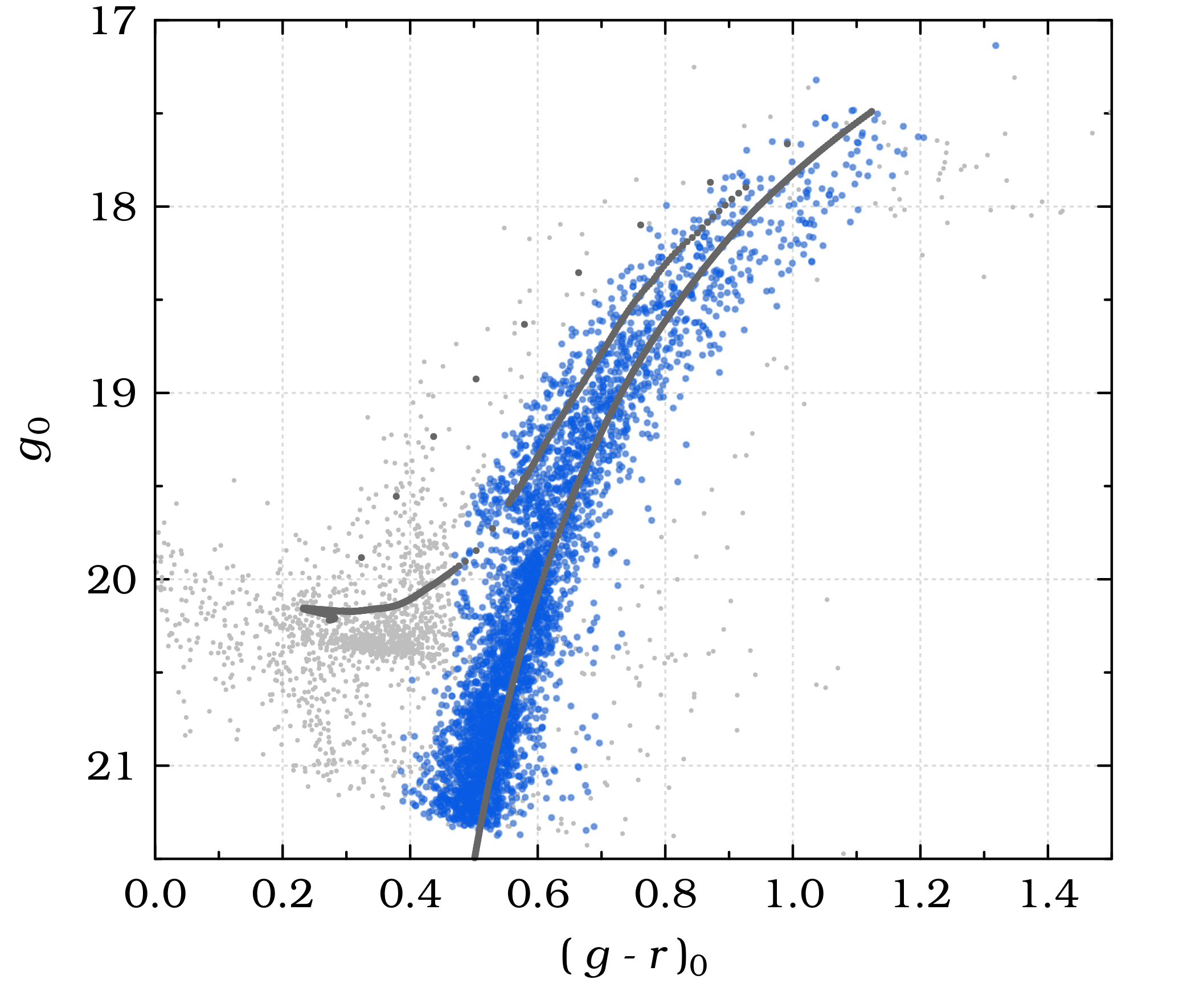}
  \end{minipage}
  \begin{minipage}[b]{\columnwidth}
    \centering
    \includegraphics[width=\columnwidth]{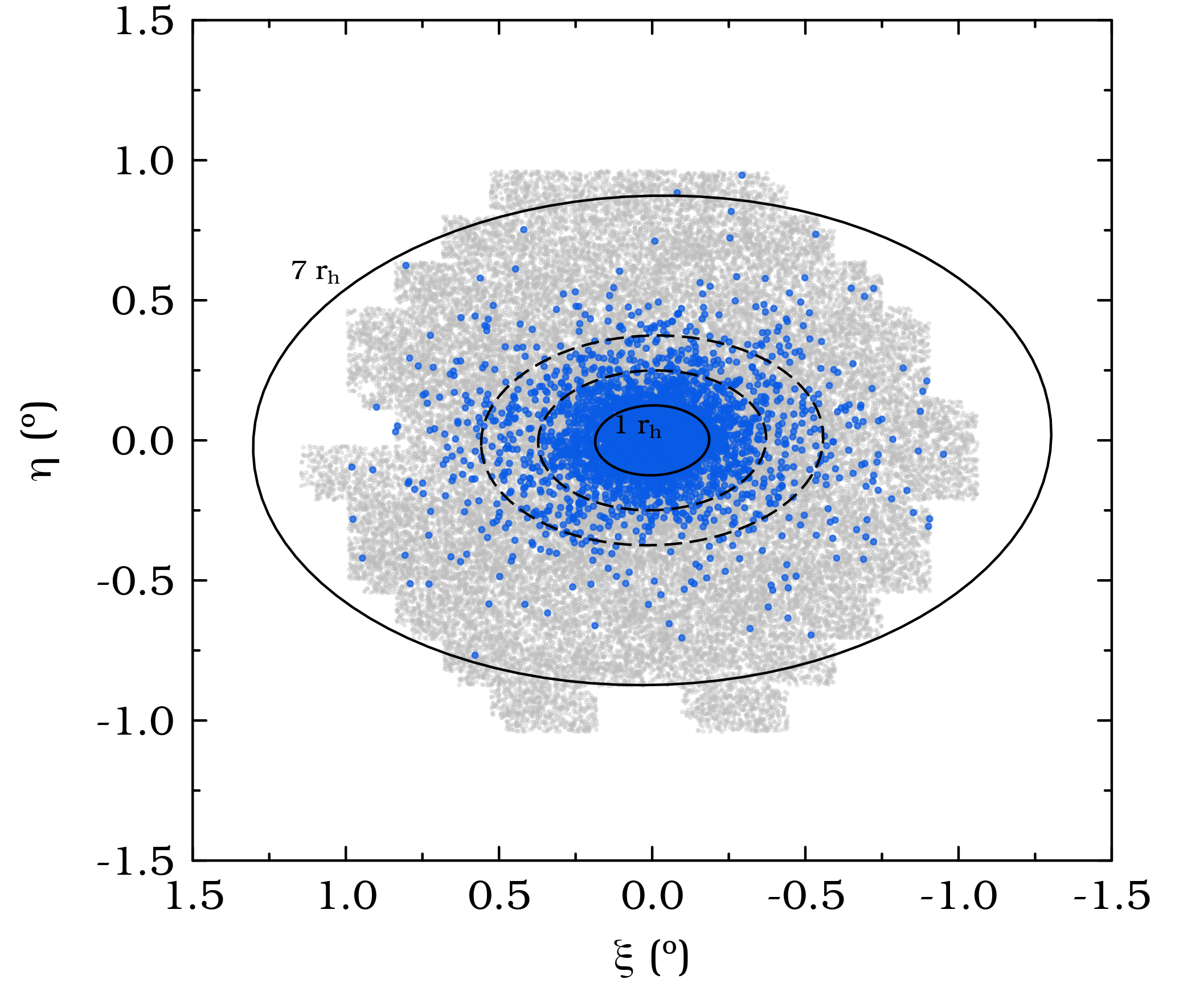}
  \end{minipage}%
  \caption{Left panel: color-magnitude diagram with DECam \textit{g}, \textit{r} photometry of selected members (blue) and the complete field observed by our DECam pointings (grey) around the Scl dSph. A BaSTI isochrone with [Fe/H] = -1.8 and 12 Gyr is overlaid. Right panel: Spatial distribution of selected members (blue) and the complete field (grey). The inner ellipse indicates the half-light radius for the Scl dSph galaxy ($\rm r_h = 11.17'$). Dashed ellipses mark two and three half-light radii. The outer ellipse covers seven half-light radii, showing the extent of our data. The structural parameters used were obtained by \cite{Munoz2018} assuming a Plummer profile.}
  \label{fig:sp_cmd}
\end{figure*}

To ensure we are analyzing the most reliable photometric metallicities, we exclude stars with final uncertainties $\sigma_{\rm [Fe/H]} > 0.5$, which also removes those with extrapolated values in the color-color space. Only stars in the RGB region are analyzed, since the MAGIC photometric depth does not reach the turn-off region of the Scl dSph and the horizontal-branch stars are not covered by the TSLTE grid. 
For our isochrone selection of candidate members, we used \textit{g}- and \textit{r}-bands from DELVE DR2, defining an interval of 0.16 in $(g - r)_0$ color around a 12 Gyr BaSTI isochrone \citep{basti} with $\rm[Fe/H] = -1.8$, the mean metallicity presented in \cite{Tolstoy2023}. This interval was chosen to encompass most of the RGB stars, recovering members from \citetalias{Tolstoy2023}. To eliminate part of the red horizontal branch within the isochrone selection, we made an additional cut, excluding stars with $(g - r)_0 < 0.47$ and $g_0 < 20.5$. 
Additionally, we verified the presence of variable stars in the sample and excluded them from further analysis, by cross-matching our catalog with the Gaia DR3 variable source catalog \citep{gaiadr3}. 
Likely galaxies are also removed using the \texttt{extended_class_g} $< 2$ criterion, as described in the DELVE DR2 paper \citep{Drlica2022}. Applying the described set of criteria, we found a sample of 3883 probable Scl dSph members. The left panel of Figure \ref{fig:sp_cmd} highlights the final sample in the color-magnitude diagram.

The spatial distribution of the selected members is presented in the right panel of Figure \ref{fig:sp_cmd}, overlapped with the field observed by the single MAGIC pointing with DECam.  
Recent studies (e.g., \citealt{Sestito2023}) have argued that the Scl dSph stellar population may extend up to more than $\sim 9 \ \rm r_h$ ($\sim 1.7\degree$). Future MAGIC data can be used to investigate these extents in the Scl dSph. However, our current coverage ($\sim 7 \, \rm r_h$; $\sim 1.3\degree$) ought to be sufficient to capture the vast majority of the spatial extent of the Scl dSph and explore its metallicity behavior.

\section{Metallicity validation}
\label{sec:comp}

Here, we will test the CaHK metallicities derived from the MAGIC data in the Scl dSph relative to previous spectroscopic metallicity studies. We will focus our discussion on the two largest spectroscopic samples available (\citealt{delosReyes22}, and \citealt{Tolstoy2023}). 
In the \hyperref[appen]{Appendix}, we present a summary of the comparison with other studies available in the literature (\citealt{Kirby2009}, \citealt{Chiti2018}, \citealt{Hill2019}, \citealt{Reichert2020}, \citealt{Walker2023}). 

\subsection{Comparison to Tolstoy et al. and de los Reyes et al.}
\label{spec}

\citet[][hereafter \citetalias{Tolstoy2023}]{Tolstoy2023} presented metallicities for more than a thousand confirmed members derived from the Ca II triplet lines contained in spectra that were homogeneously acquired with the VLT Fibre Large Array Multi-Element Spectrograph (FLAMES) low-resolution data. 
Following the suggested quality flags described in that paper, only members in their study with signal-to-noise ratio $> 13$ and no flag indicating sky subtraction residuals were selected. This gives a sample of 1148 stars observed and measured by MAGIC in common with \citetalias{Tolstoy2023}.

The other spectroscopic study we compare to, \citet[][hereafter \citetalias{delosReyes22}]{delosReyes22}, compiled data from previous works obtained with the Deep Imaging Multi-Object Spectrometer (DEIMOS) on the Keck II telescope \citep{deimos} and by the Dwarf Abundances and Radial Velocities Team (DART; \citealt{dart}). 
Metallicities and $\alpha$-abundances were derived using spectral synthesis by \cite{Kirby2010}, being both an independent dataset and method compared to the \citetalias{Tolstoy2023} analysis. From the 465 stars in their sample, we found 366 in common with the MAGIC survey, located mainly in the central region ($\sim1 \  r_h$) of the Scl dSph galaxy.

\begin{figure*}[pt!]
\centering
  \begin{minipage}[b]{\columnwidth}
    \centering
    \includegraphics[width=\columnwidth]{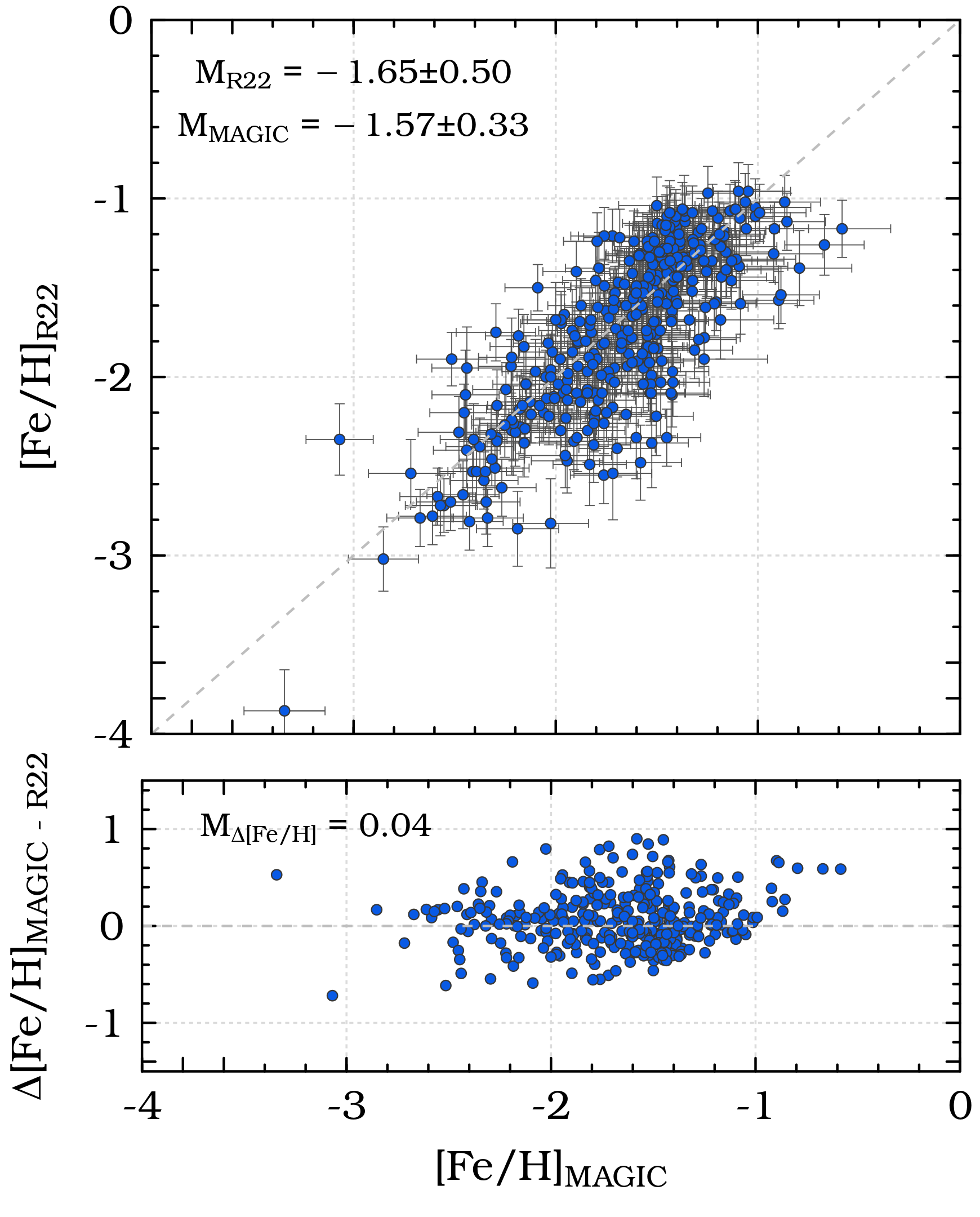}
  \end{minipage}
  \begin{minipage}[b]{\columnwidth}
    \centering
    \includegraphics[width=\columnwidth]{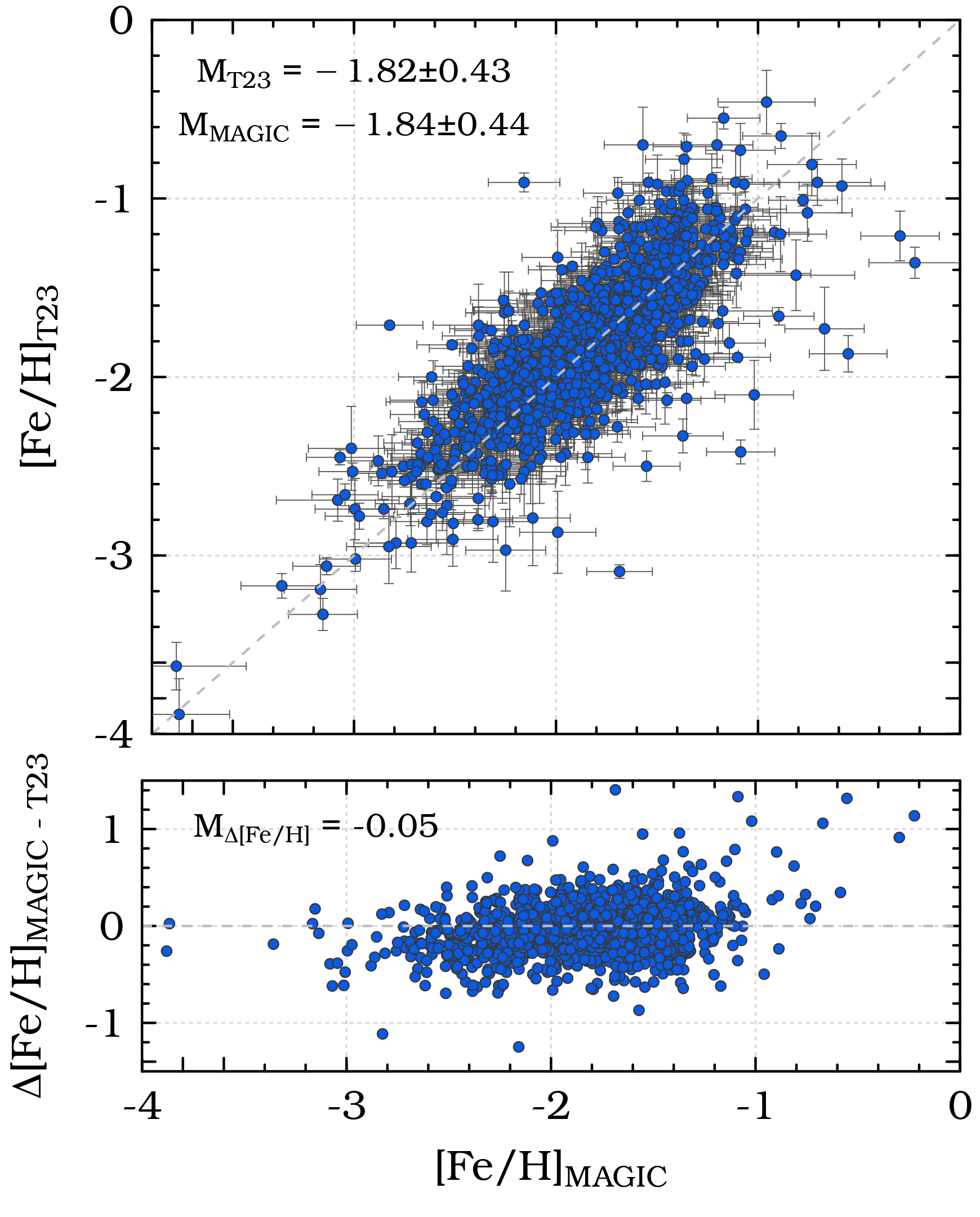}
  \end{minipage}%
  \caption{Top panels: comparison between the derived photometric metallicity values in this work and in spectroscopic work in the literature by \cite{delosReyes22} (left) and \cite{Tolstoy2023} (right). Median values and median absolute deviations ($\rm M_{MAGIC}, \ M_{R22}, \ and \ M_{T23}$) are indicated in the top left corner of each panel. Bottom panels: difference between the $\rm[Fe/H]_{MAGIC}$ and the spectroscopic estimates. $\rm M_{\Delta[Fe/H]}$ indicates the median value.}
  \label{fig:feh_comp}
\end{figure*}

The upper panels in Figure \ref{fig:feh_comp} show the comparison between the MAGIC-derived photometric metallicities presented in this work and the spectroscopic values reported by \citetalias{delosReyes22} and \citetalias{Tolstoy2023}. The lower panels present the difference between the MAGIC estimates and the spectroscopic data, with which the median offsets were calculated. Both estimates exhibit an overall good agreement with the $\rm[Fe/H]_{MAGIC}$, with no notable trend compared to the larger \citetalias{Tolstoy2023} sample.

For $\rm[Fe/H]_{MAGIC} < -2.5$, we do not have a good sampling in \citetalias{delosReyes22}, since it mainly covers the central region of Scl dSph, which is more metal-rich than its outskirts \citep{Taibi2022}. In the same interval, we note that the 73 stars in common with \citetalias{Tolstoy2023} tend to be more metal-poor in the MAGIC data, being $\sim 0.18$ dex lower than the reported spectroscopy. Stars with $\rm[Fe/H]_{MAGIC} > -1$ in both panels tend to be overestimated by $\sim 0.49$ dex comparing to \citetalias{delosReyes22}, and $\sim 0.32$ dex to \citetalias{Tolstoy2023}. 
Between $\rm[Fe/H]_{MAGIC} = -2.5$ and $\rm[Fe/H]_{MAGIC} = -1$, the photometric values are in excellent agreement with the spectroscopic ones. The median offset observed in this interval for \citetalias{delosReyes22} is $0.03$ dex and for \citetalias{Tolstoy2023} is $-0.05$ dex, with median absolute deviations of $0.26$ dex and $0.22$, respectively. 
The overall median offset for \citetalias{delosReyes22} is $0.04$ dex. For \citetalias{Tolstoy2023}, we find $-0.05$ dex. 
Except for the caveats mentioned, these comparisons make us confident that MAGIC photometric [Fe/H] values are robust and can be used to study stellar populations properties of this dwarf galaxy, for the bulk of its metallicity range.
We note that the MAGIC metallicity calibration is based only on the synthetic isochrones and has not been tuned to the spectroscopic data.

\begin{figure}[t]
\begin{center}
\includegraphics[width=\columnwidth]{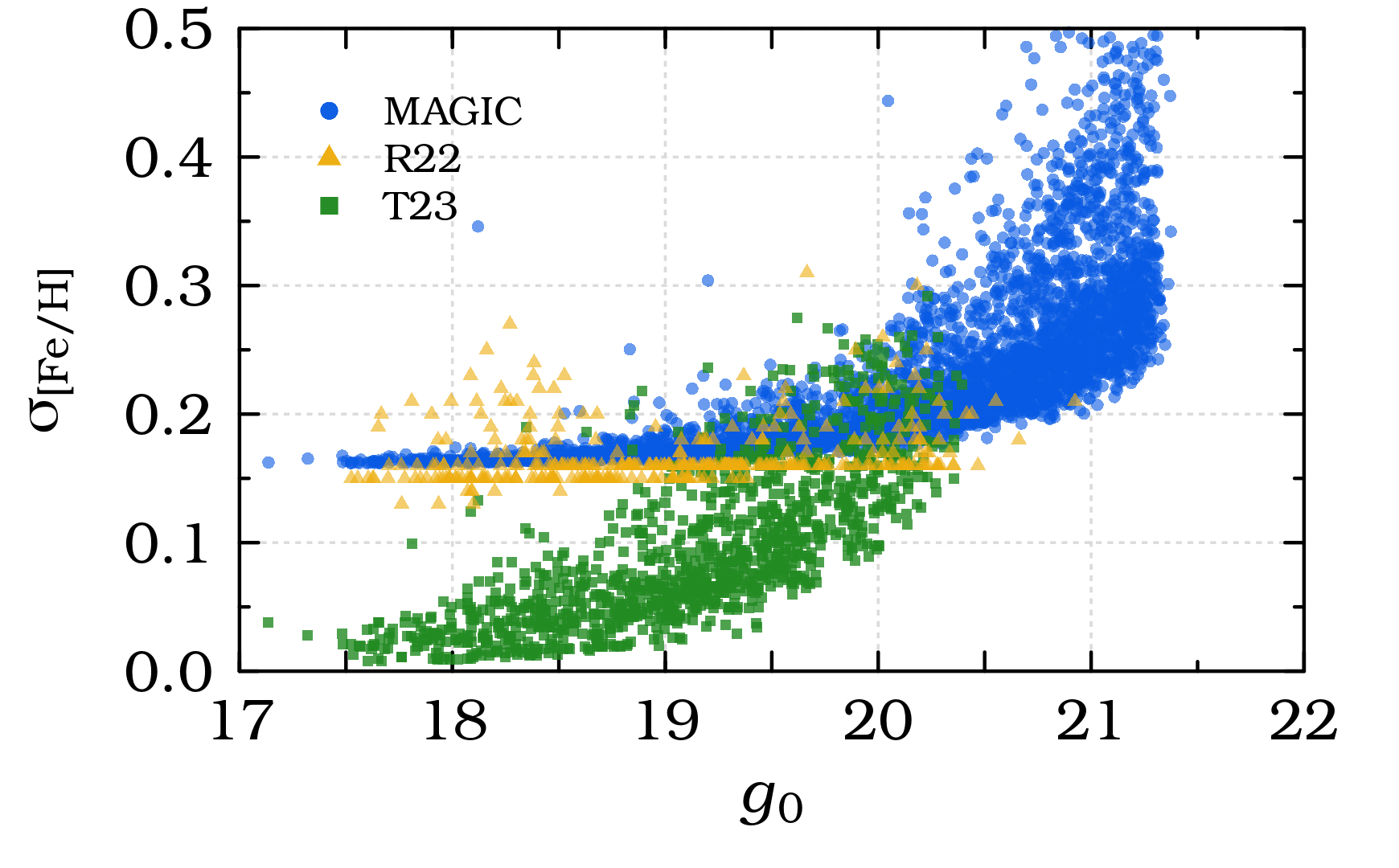}
\includegraphics[width=\columnwidth]{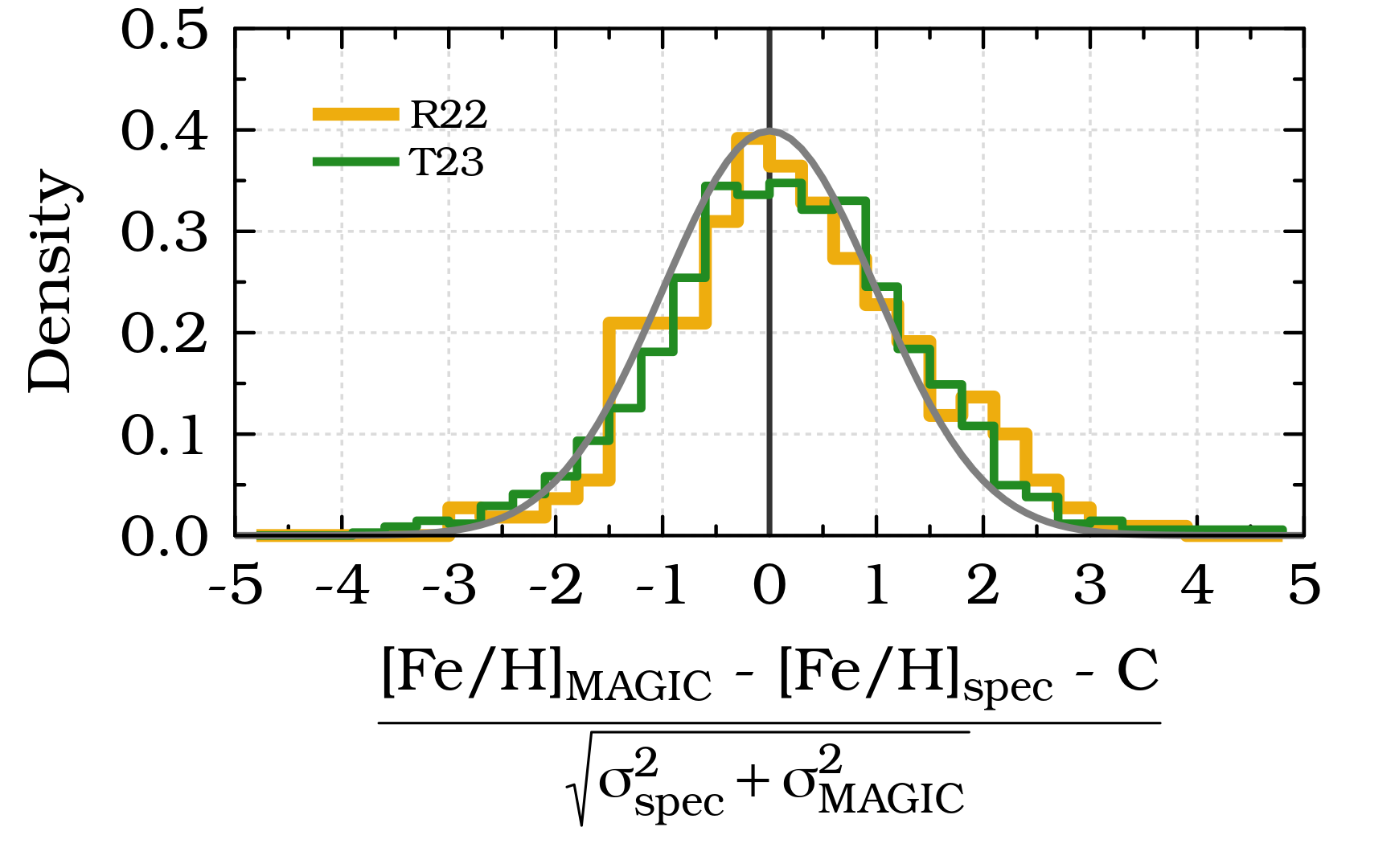}
\caption{Top panel: metallicity uncertainties reported by the original works as a function of $g_0$ magnitude for members observed by the MAGIC survey (blue points), and stars in common with \citeauthor{delosReyes22} (\citeyear{delosReyes22}; R22 -- yellow triangles) and with \citeauthor{Tolstoy2023} (\citeyear{Tolstoy2023}; T23 -- green squares). Bottom panel: normalized difference histograms comparing MAGIC estimates with \citetalias{delosReyes22} (yellow thick line) and \citetalias{Tolstoy2023} (green thin line). The gray curve illustrates a Gaussian function with mean $=0$ and standard deviation $=1$ for comparison.}
\label{fig:norm_hist}
\end{center}
\end{figure}

The metallicity uncertainties ($\sigma_{\rm[Fe/H]}$) for all members of the Scl dSph observed by MAGIC, as well as for the stars in common between MAGIC and spectroscopic works, as a function of $g_0$ magnitude are presented in the upper panel of Figure \ref{fig:norm_hist}. Note that our uncertainties are usually larger than the values reported in \citetalias{delosReyes22} and \citetalias{Tolstoy2023} due to the systematic uncertainty floor adopted. For stars with $g_0 \sim 20$, our uncertainties are comparable to those reported by \citetalias{Tolstoy2023}. Nearly all stars with $g_0 < 20$ have associated $\sigma_{\rm[Fe/H]}$ smaller than $0.3$ dex, and approximately 86\% of the entire sample fall within this uncertainty threshold.

The bottom panel in Figure \ref{fig:norm_hist} presents the normalized difference histogram, i.e., 
\begin{equation}
\frac{\rm [Fe/H]_{MAGIC} - [Fe/H]_{spec} - C}{\sqrt{\sigma^2_{\rm spec} + \sigma_{\rm MAGIC}^2}},    
\end{equation} \\
where $\rm C$ is the overall median offset between MAGIC metallicity estimates and the spectroscopic works, and $\sigma$ are the respective metallicity uncertainties. The yellow line corresponds to the comparison with \citetalias{delosReyes22}'s work, and the green line, to \citetalias{Tolstoy2023}'s work. Means, medians, standard deviations and median absolute deviations for both distributions are presented in Table \ref{main_tab}. As previously noted in Figure \ref{fig:feh_comp}, there is a good agreement between MAGIC photometric metallicities and the spectroscopic estimates, with standard deviations of both distributions close to 1. The gray curve represents a Gaussian curve with mean $\mu = 0$ and standard deviation $\rm SD = 1$, serving as a reference to highlight the agreement between the estimates.

\begin{figure}[t]
\begin{center}
\includegraphics[width=\columnwidth]{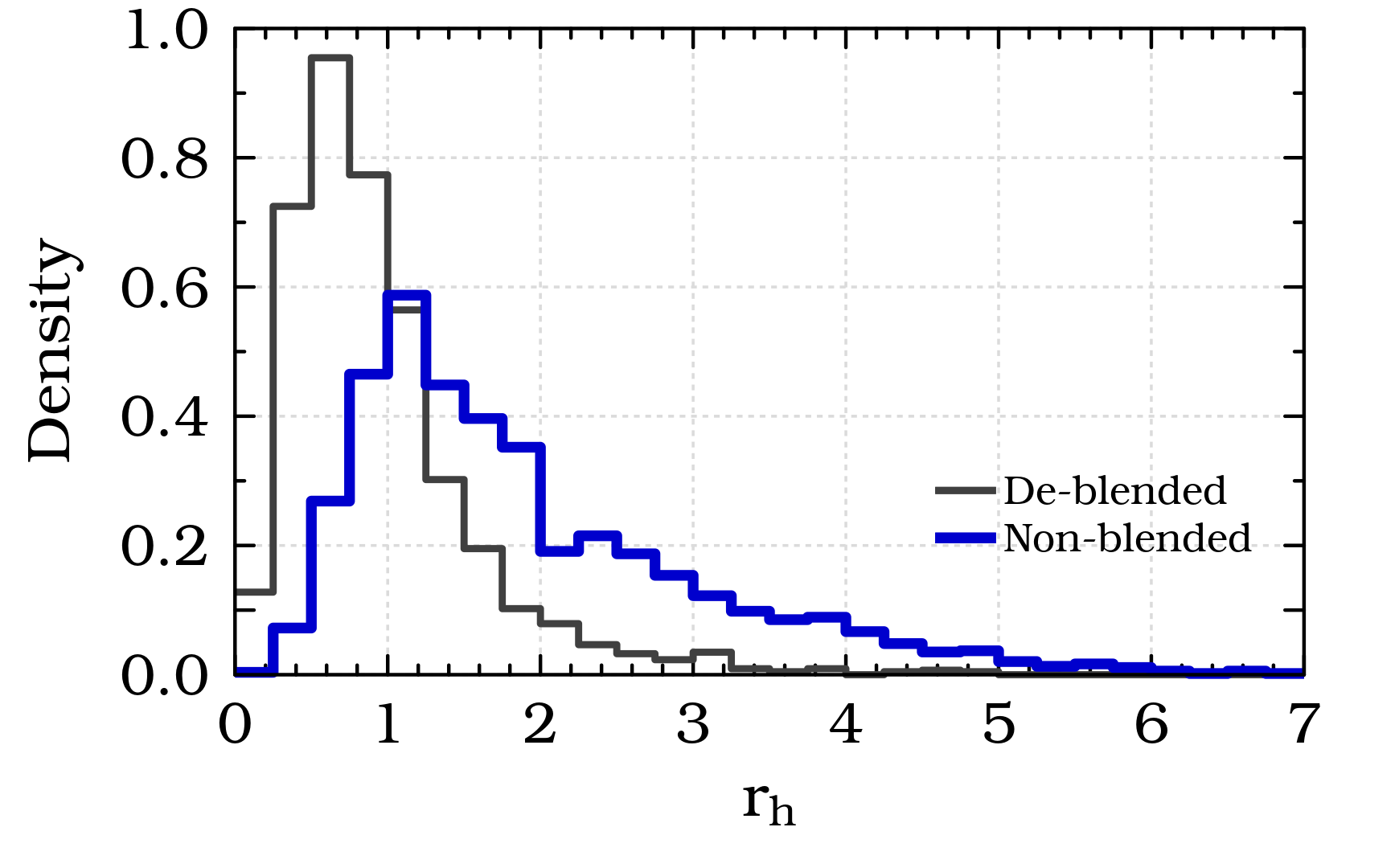}
\includegraphics[width=\columnwidth]{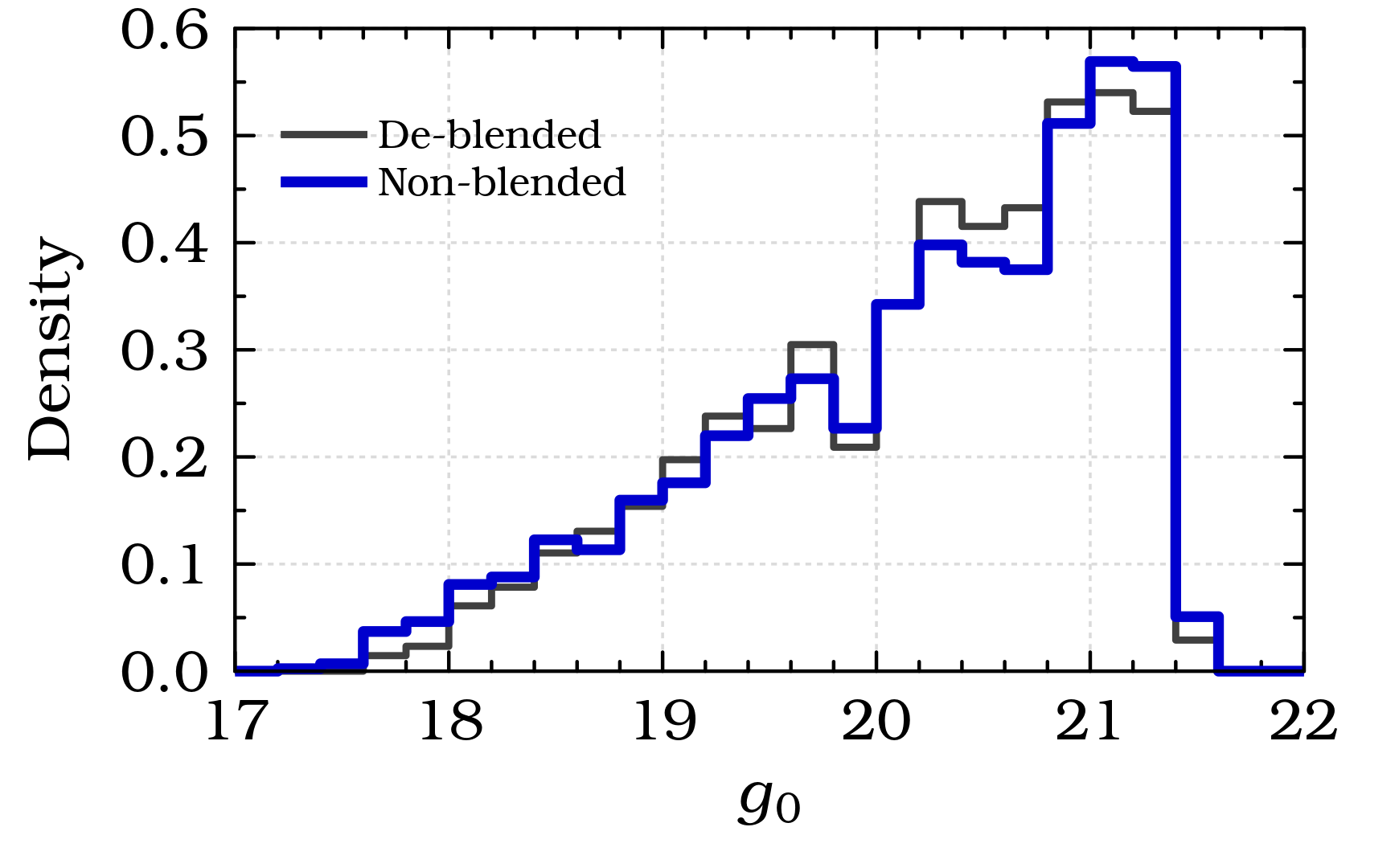}
\caption{Radial (top panel) and $g_0$ magnitude distribution (bottom panel) for stars de-blended (blue thick line) and non-deblended stars (grey thin line). The radial distribution is presented in units of half-light radius, using $\rm r_h = 11.17'$ for the Scl dSph galaxy (\citealt{Munoz2018}).}
\label{fig:debl}
\end{center}
\end{figure}

As discussed in Section \ref{magic}, stars with photometry that has been de-blended in either the CaHK, \textit{g}-, or \textit{i}-band data receive an indicator in our catalogue (\texttt{feh_flag_phot $= \{2,3\}$}), following standard \texttt{Source Extractor} and \texttt{PSFEx} flags. Nearly 44\% of the selected members in the Scl dSph were successfully de-blended by \texttt{PSFEx}. The top panel of Figure \ref{fig:debl} presents the radial distribution for stars de-blended (blue thick line) comparing with those that did note require not de-blending (grey thin line). It is evident that de-blended stars are mainly found in the inner region, while not de-blended stars populate all radii in the galaxy. The bottom panel shows the magnitude distribution in $g$ for both subsamples, showing no dependence of the de-blended sources with brightness.

\begin{figure}[t]
\begin{center}
\includegraphics[width=\columnwidth, trim={0 2cm 0 0},clip]{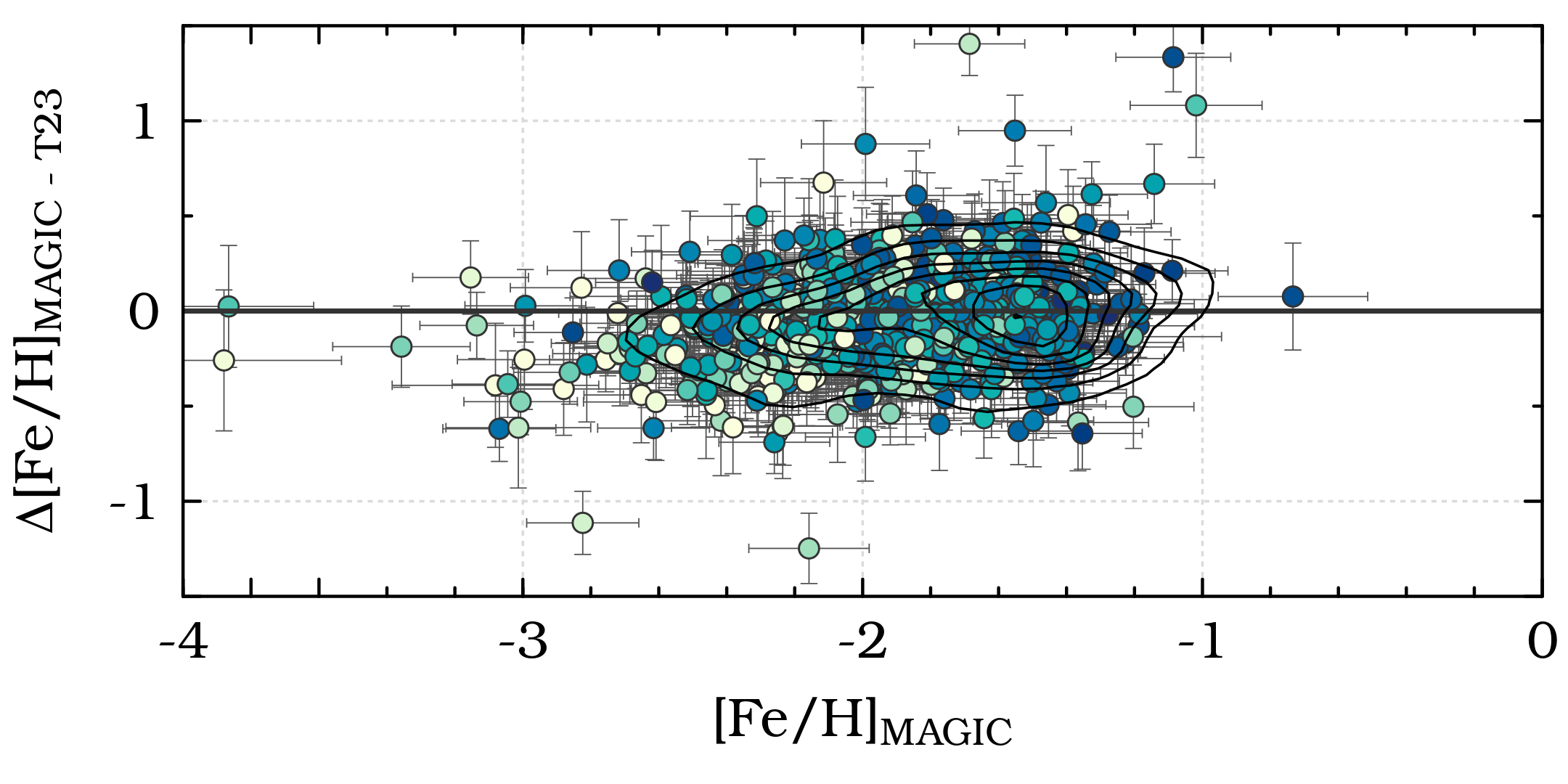}
\includegraphics[width=\columnwidth]{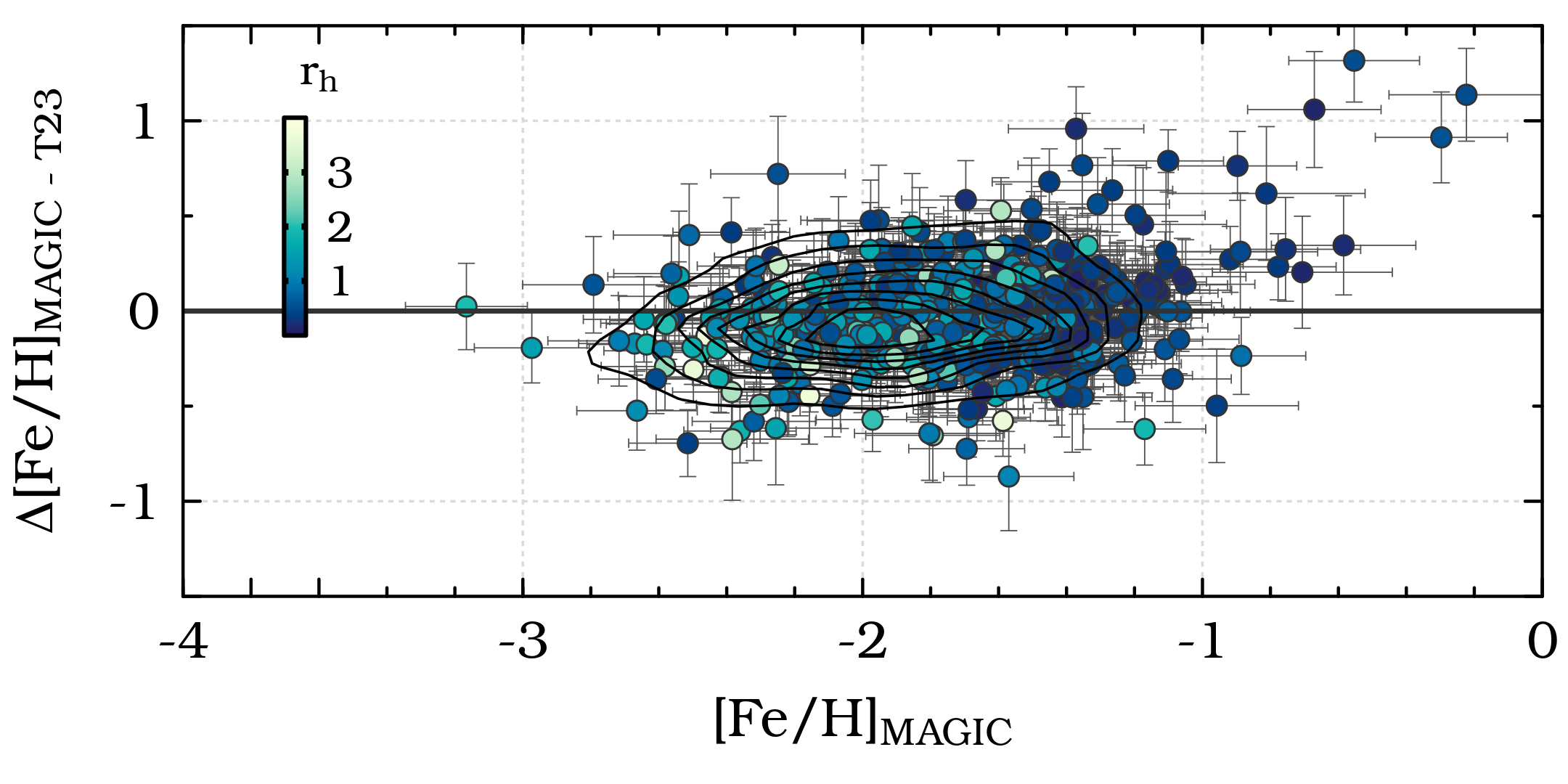}
\caption{Comparison of photometric metallicities and spectroscopic metallicities from \citetalias{Tolstoy2023} separating the sample between stars with (bottom panel) and without (top panel) de-blended photometry. 
Contours corresponding to non-blended sources are plotted over the de-blended sample and vice versa. Colorbar indicates the radial distance.}
\label{fig:flag}
\end{center}
\end{figure}

We examine the metallicity performance for sources that are/are not de-blended, and confirm that the de-blending procedure by \texttt{PSFEx} does not significantly affect the resulting photometric metallicity. Figure \ref{fig:flag} presents the comparison with \citetalias{Tolstoy2023} for stars that did and did not need de-blending in the photometry. 
Stars that did not require de-blending are shown in the upper panel, while stars that were de-blended are plotted in the lower panel. Each plot shows the density contours corresponding to the complementary sample. The radial distance is indicate by the color, with all sources beyond $4 \ r_h$ colored with yellow.
When comparing with \citetalias{Tolstoy2023} data, we found 484 members that are de-blended. This process does not have a high impact on our metallicity estimates, since both subsets present a small offset ($-0.03$ for de-blended sources, and $-0.07$ for non-blended). 

\subsection{Photometric metallicities for carbon-enhanced metal-poor stars}
\label{sec:cfe}

It is important to verify the photometric metallicity estimates for carbon enhanced metal-poor (CEMP; $\rm [Fe/H] < -1 \ and \ [C/Fe] \gtrsim +0.7$) stars \citep{Beers2005}, since high carbon-abundances increase the strength of the CN feature blueward of the Ca II K line, suppressing the flux in the CaHK filter and making the star appear more metal-rich (\citealt{Yoon2020}, \citealt{Placco2025}). 
Therefore, an additional analysis was performed considering the \cite{Chiti2018} sample, which obtained medium-resolution spectra with the Michigan/Magellan Fiber System on the Magellan-Clay telescope for 100 stars in the Scl dSph. They estimated [Fe/H], from the Ca II K line, and [C/Fe] from the CH G-band at $\sim 430 \rm \ nm$. Here, we found 47 stars in common between \citetalias{Tolstoy2023}, \cite{Chiti2018} and our survey, with which we can compare the photometric estimates and observe the influence of the carbon abundance on the photometric metallicity determination.

\begin{figure}[t]
\begin{center}
\includegraphics[width=\columnwidth]{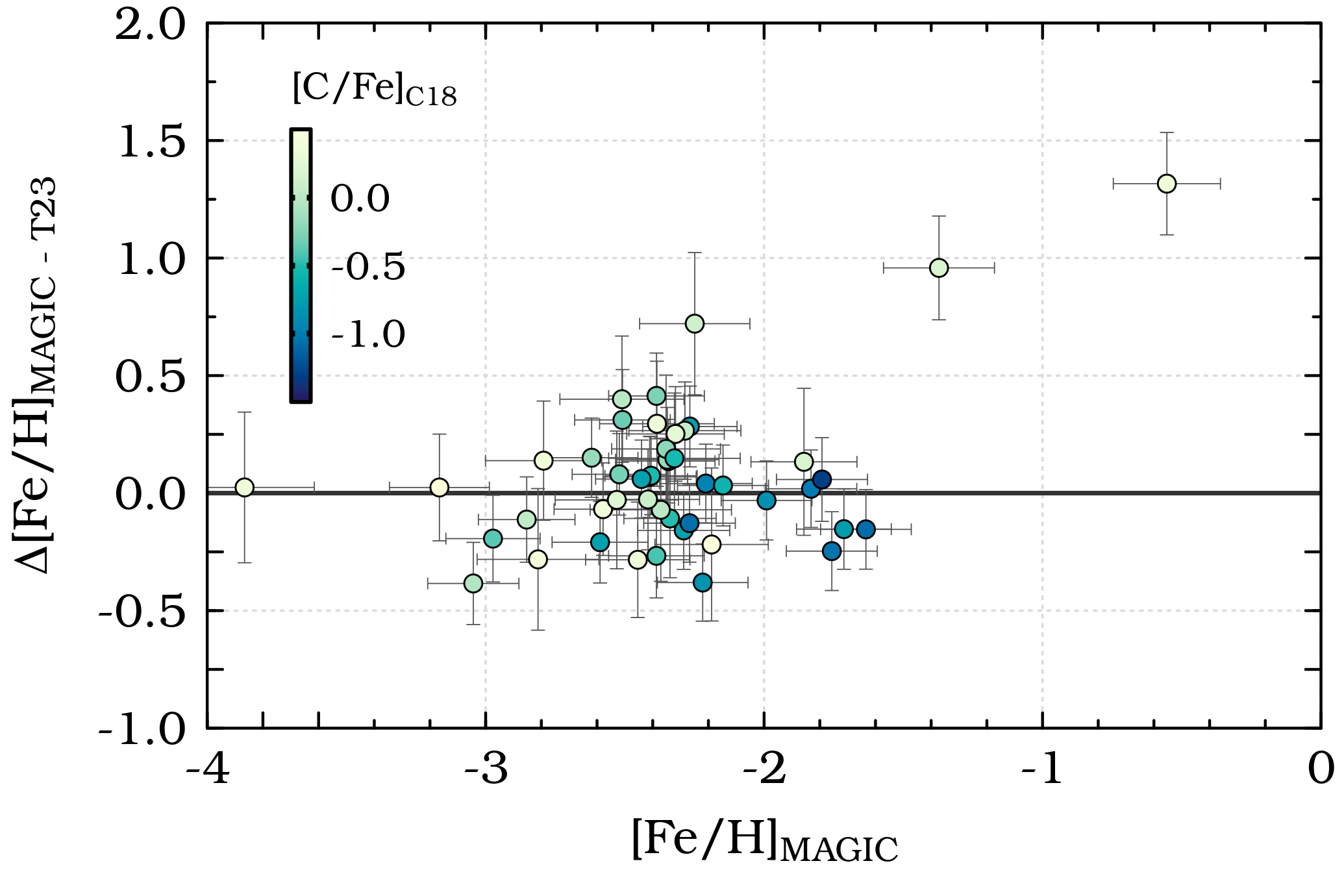}
\caption{Difference between photometric metallicity and spectroscopic values, colored according to the measured carbon abundance for each star. Spectroscopic [Fe/H] obtained by \cite{Tolstoy2023} and [C/Fe] are reported by \cite{Chiti2018}.}
\label{fig:cfe}
\end{center}
\end{figure}

Figure \ref{fig:cfe} shows the difference between $\rm[Fe/H]_{MAGIC}$ and $\rm[Fe/H]_{T23}$, colored by the carbon abundance reported in \cite{Chiti2018}, where the carbon-to-iron ratio are not corrected for the evolutionary state of the star \citep{Placco2014}. The subsample exhibits a median offset of $0.03$ dex. Three stars present values of $\Delta_{\rm[Fe/H]_{MAGIC - T23}} > 0.5$ and have $\rm [C/Fe]_{C18} > +0.1$, which could indicate an effect of the CN band absorption found at the edge of bluer wavelengths of the filter. 
This effect is also noted in, e.g., \cite{Martin2024} and \cite{Placco2025}, and could possibly explain the overestimation of [Fe/H] observed in Figure \ref{fig:feh_comp} for stars with $\rm[Fe/H]_{MAGIC} > -1$. 
However, the small number of stars following this trend limits the inspection of a possible influence of carbon abundances in the metallicity overestimation in that regime.

Nevertheless, we do not expect this systematic to meaningfully affect the analysis presented in this paper. 
General evidence that this systematic ought not to affect our interpretation follows from the comparisons in Figure \ref{fig:feh_comp}.
The MAGIC photometric metallicities show good agreement with spectroscopic metallicities across large numbers of stars in the Scl dSph that were not pre-selected based on carbon, likely spanning the characteristic range of [C/Fe] for this galaxy.

\section{Results \& Discussion}
\label{sec:res}

After verifying the performance of the photometric metallicities, we can inspect the characteristics of the Scl dSph. The new sample offers a significant advantage over spectroscopic ones, as it is both spatially unbiased and represents the largest dataset of metallicities in the literature to date ($\sim 3800$ stars) in any dSph galaxy. This is the first application of the MAGIC data, showing how our photometric metallicities can help construct a more complete spatial view of systems as distant as the Milky Way classical satellites. 

\subsection{Metallicity distribution}
\label{mdf}

\begin{figure*}[t]
\begin{center}
\tikzmark{a}\includegraphics[width=\columnwidth]{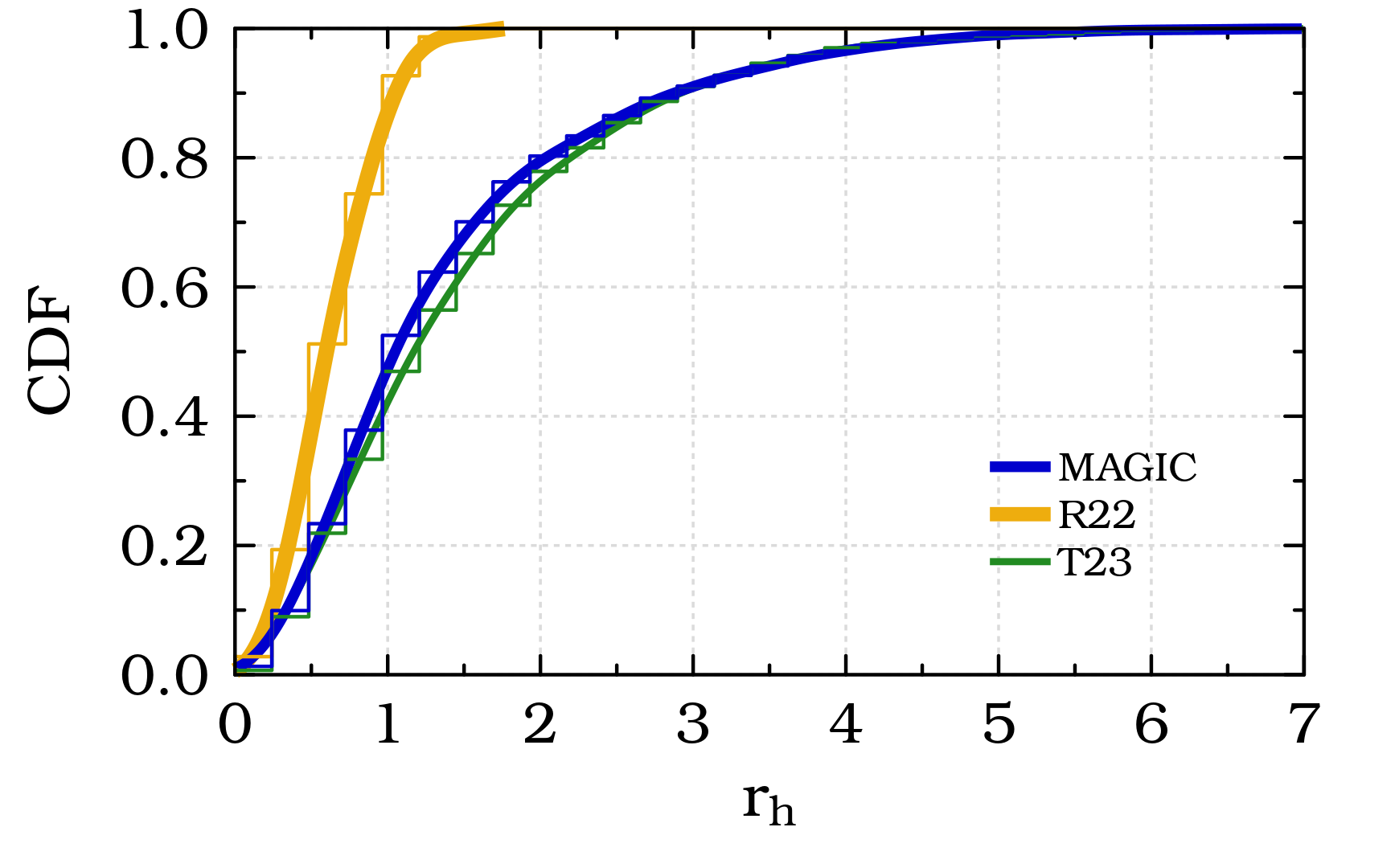}
\tikzmark{b}\includegraphics[width=\columnwidth]{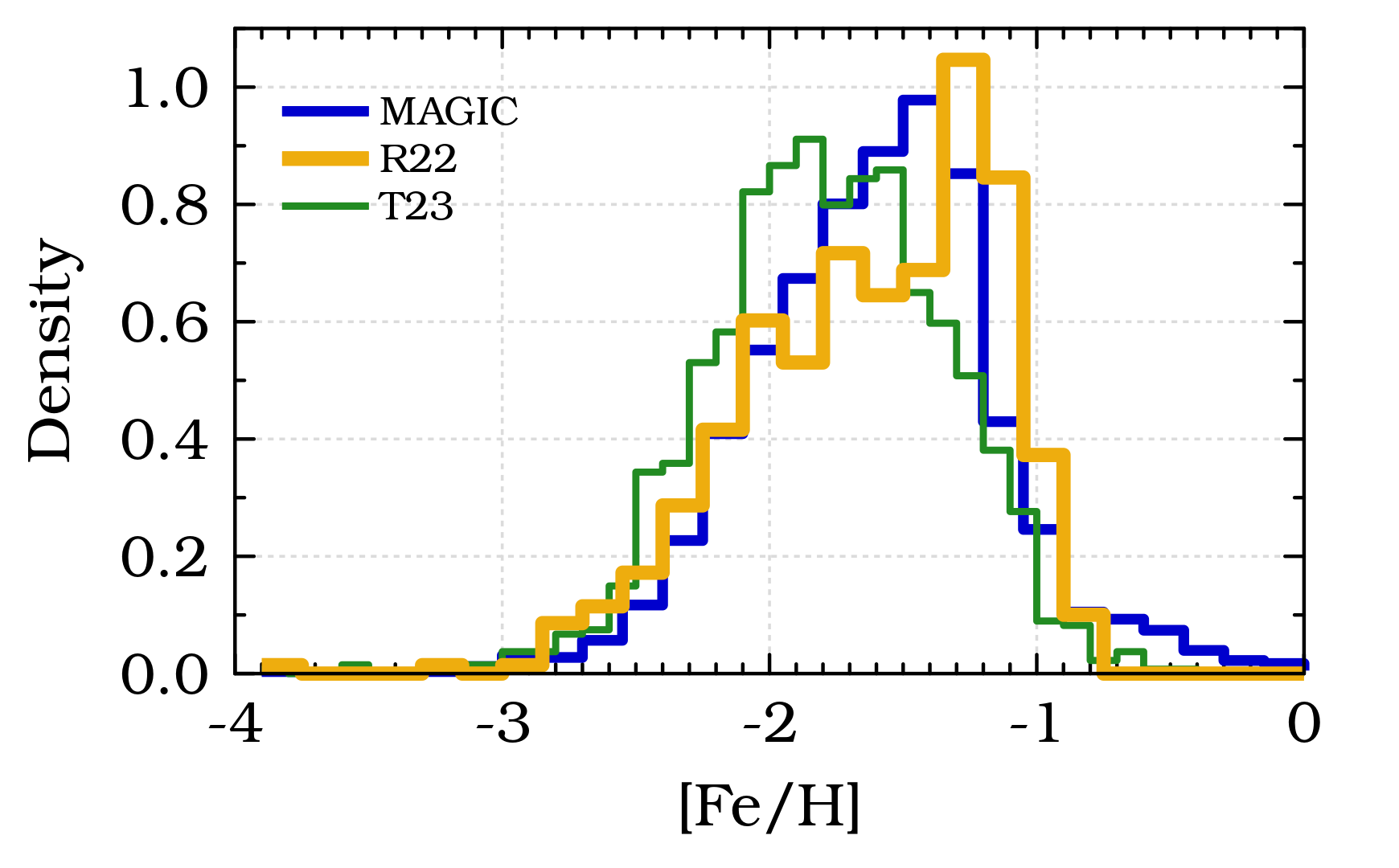}
\tikzmark{c}\includegraphics[width=\columnwidth]{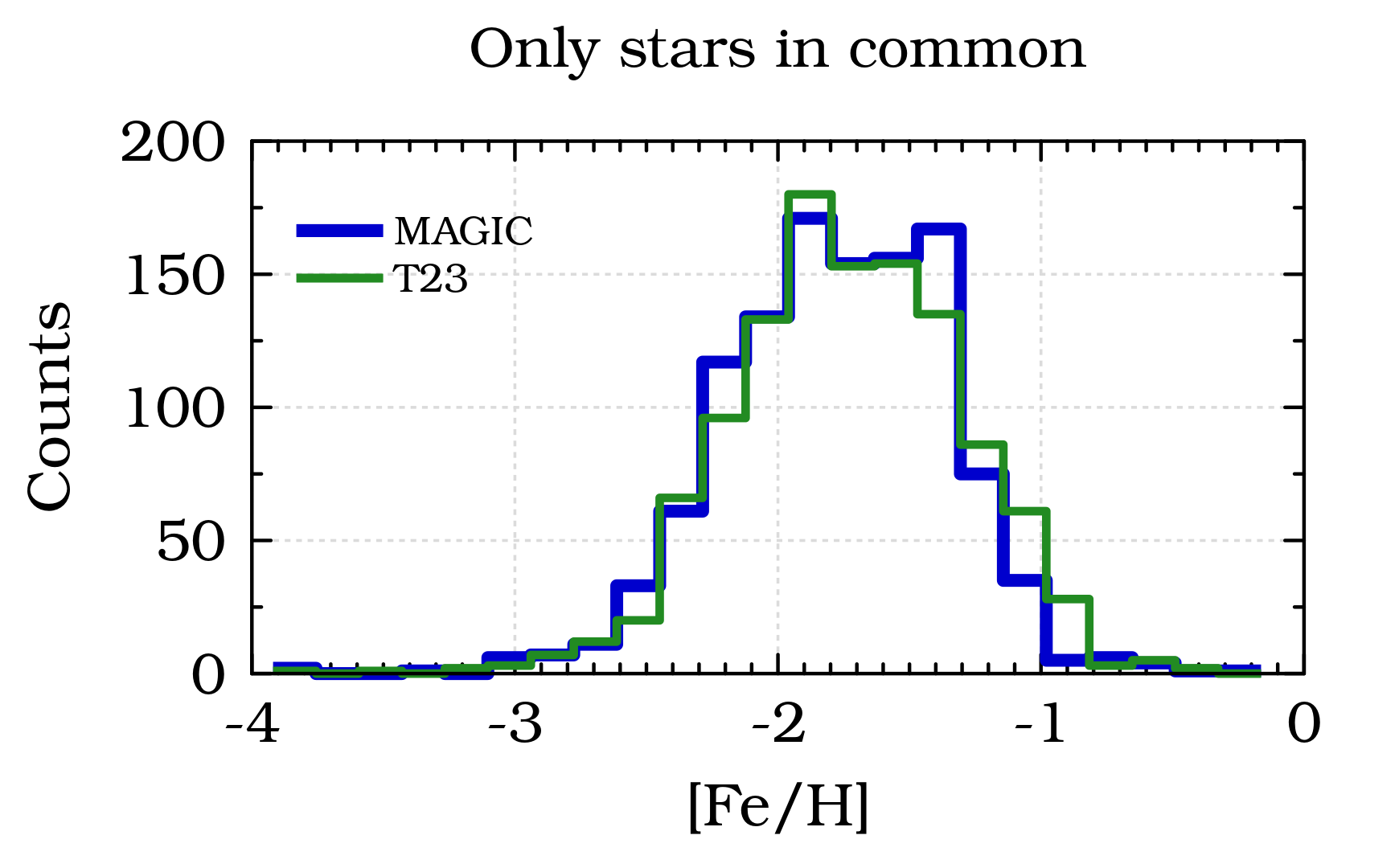}
\tikzmark{d}\includegraphics[width=\columnwidth]{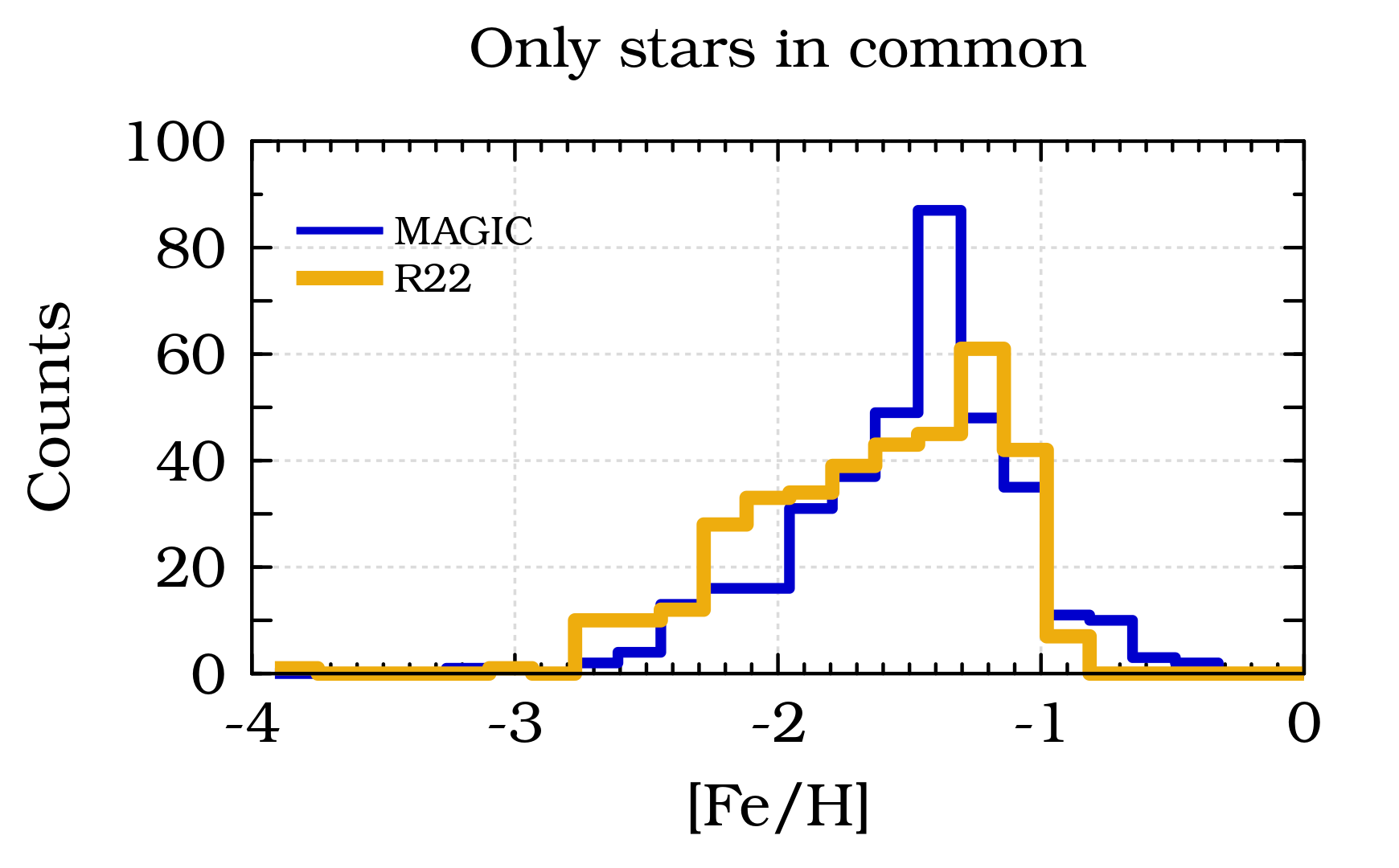}
\caption{Panel (a): cumulative distribution function as a function of half-light radius. The blue line indicates the data obtained with MAGIC observations; the yellow thick line, with \citeauthor{delosReyes22} (\citeyear{delosReyes22}; R22); the green thin line, with \citeauthor{Tolstoy2023} (\citeyear{Tolstoy2023}; T23). Panel (b): comparison of the MDF for the different studies of the Scl dSph galaxy. Panel (c): comparison between photometric (MAGIC) and spectroscopic (T23) data for members in common. Panel (d): comparison between photometric (MAGIC) and spectroscopic (T23) data for members in common.}
\label{fig:mdf}
\InsertLabels{7.3cm}{-4.5cm}
\end{center}
\end{figure*}

Panel (a) of Figure \ref{fig:mdf} presents the cumulative distribution function of the projected distance measure in $\rm r_{h}$ of the stars in the Scl dSph observed by \citetalias{delosReyes22}, \citetalias{Tolstoy2023}, and this study, indicating the spatial distribution of the members. 
The data used in \citetalias{delosReyes22} (yellow thick line) are more restricted to the core, which is the more metal-rich part of the galaxy, with a coverage of $\sim 1.8 \ \rm r_h$. Both MAGIC and \citetalias{Tolstoy2023} data cover out to $\sim 7 \ \rm r_h$.

Panel (b) of Figure \ref{fig:mdf} shows the different MDFs for each work. The \citetalias{delosReyes22} MDF peaks at higher values, with a median of $\rm[Fe/H]_{R22} = -1.62$, and exhibits an extended metal-poor tail. On the other hand, \citetalias{Tolstoy2023} (green thin line) covered a larger area and observed more metal-poor stars toward the outskirts of the galaxy, resulting in a median $\rm[Fe/H]_{T23} = -1.83$. 
The spatial coverage obtained in this work spans the inner and outer regions of the galaxy, since there is no prior target selection for our observations, as required for spectroscopy. Therefore, our data displays an MDF with a peak more metal-rich than \citetalias{Tolstoy2023}, more consistent with \citetalias{delosReyes22}.  
The main statistical parameters for the MDF with MAGIC estimates are presented in Table \ref{main_tab}. 
The median metallicity obtained is $\rm[Fe/H]_{MAGIC} = -1.65$ and a Kolmogorov-Smirnov test reinforces that the MAGIC MDF is statistically different from \citetalias{Tolstoy2023} spectroscopic sample, with a p-value $< 0.001$, while comparing with \citetalias{delosReyes22} sample we find p-value $\sim 0.10$.

Panel (c) of Figure \ref{fig:mdf} shows the MDFs of MAGIC photometric estimates (blue line) and the spectroscopic ones from \citetalias{Tolstoy2023} (green line), when considering only the members in common between the samples. 
Both MDFs appear similar, with a MAGIC median value of $\rm[Fe/H]_{MAGIC} = -1.84$ in excellent agreement with \citetalias{Tolstoy2023} measurements. This result suggests that the discrepancy observed in the MDFs of the entire MAGIC and \citetalias{Tolstoy2023} samples presented in the top panel of Figure \ref{fig:mdf} is likely due to spatial sampling. 
Panel (d) of Figure \ref{fig:mdf} presents the MDFs of MAGIC (blue line) and \citetalias{delosReyes22} spectroscopic metallicities (yellow thick line), considering only the members in common between the samples. In this case, we find a median value of $\rm[Fe/H]_{MAGIC} = -1.53$ and $\rm[Fe/H]_{R22} = -1.66$, and  different shapes, reflecting the dispersion observed in Figure \ref{fig:feh_comp}.

\begin{figure}[t]
\begin{center}
\includegraphics[width=\columnwidth, trim={0 2cm 0 0}, clip]{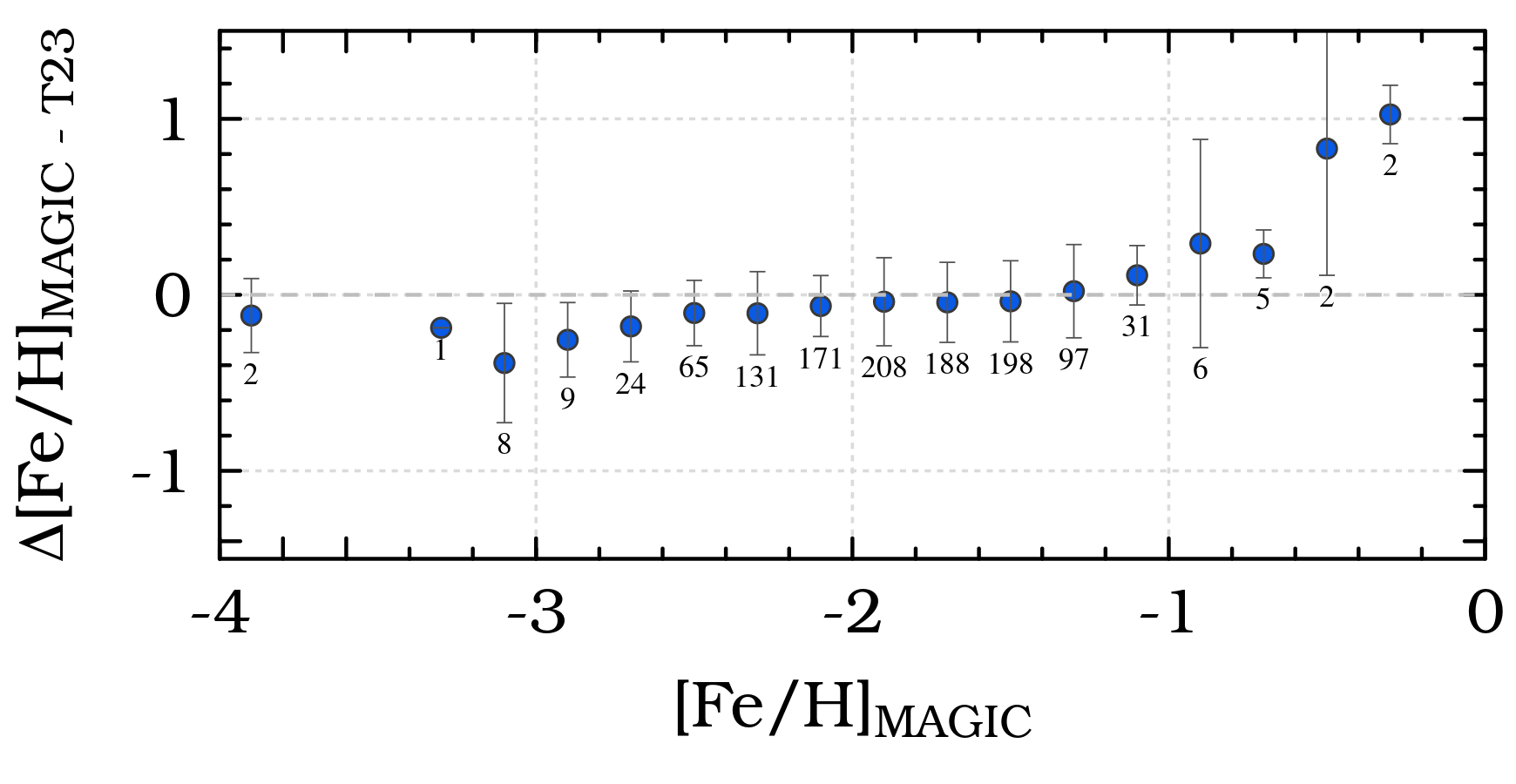}
\includegraphics[width=\columnwidth]{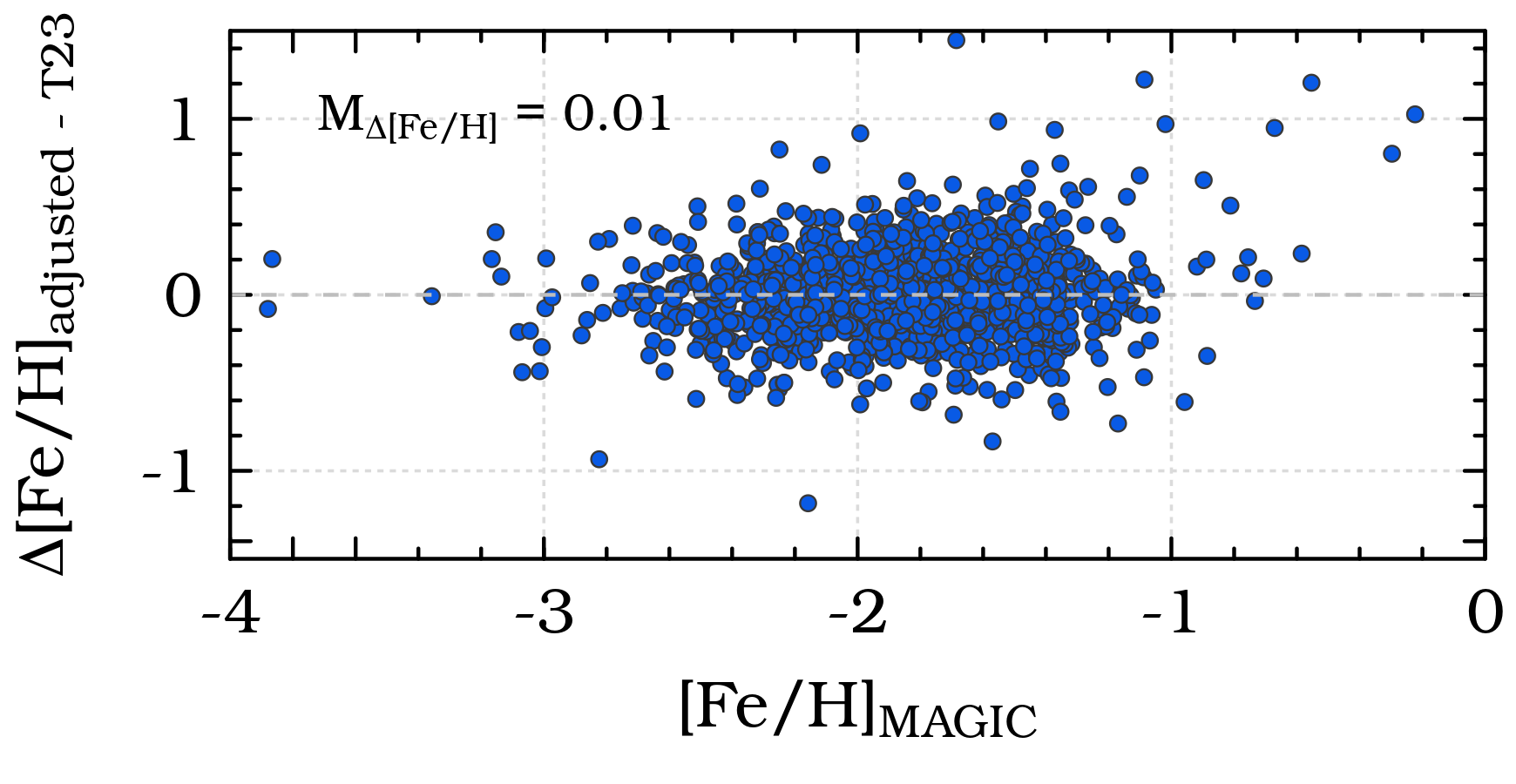}
\includegraphics[width=\columnwidth]{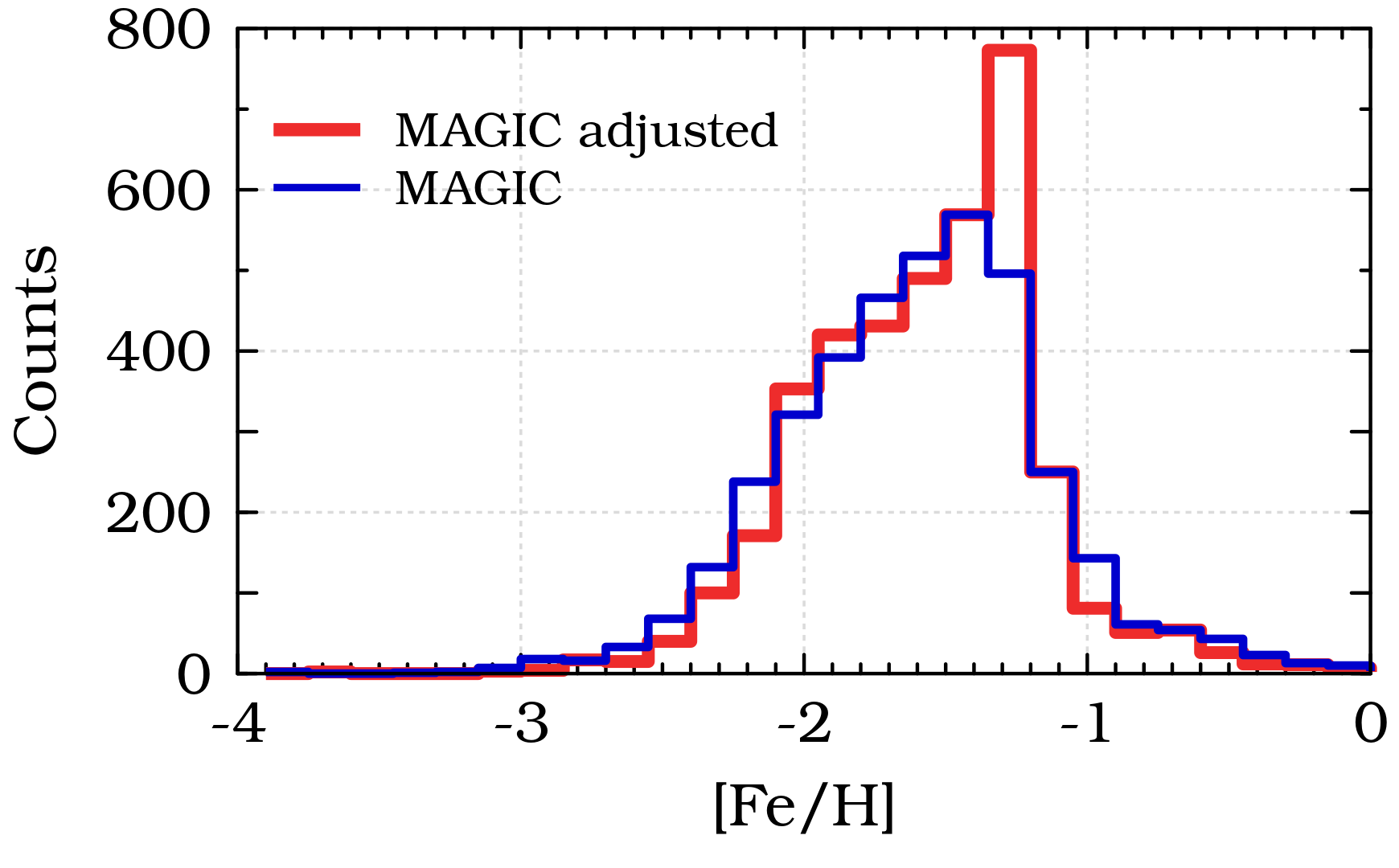}
\caption{Top panels: median difference between MAGIC and \citetalias{Tolstoy2023} estimates in each bin used to calculated $\rm[Fe/H]_{adjusted}$ and residual plot after the adjustment of $\rm[Fe/H]_{MAGIC}$. Vertical bars correspond to the median absolute deviation. The number of stars comprised in each bin is indicated. $\rm M_{\Delta[Fe/H]}$ indicates the median value. Bottom panel: MDFs for estimated MAGIC metallicities (thin blue line) and adjusted values ($\rm[Fe/H]_{adjusted}$, thick red line).}
\label{fig:adj_proc}
\end{center}
\end{figure}

We present an adjusted metallicity that accounts for systematics discussed in Section~\ref{spec}, that is the metallicity overestimation for stars with $\rm[Fe/H]_{MAGIC} > -1.0$ and the slight metallicity underestimation for $\rm[Fe/H]_{MAGIC} < -2.5$. The adjustment was performed by first dividing the data into bins of $0.2$ dex in $\rm[Fe/H]_{MAGIC}$ and estimating the median offset between $\rm[Fe/H]_{MAGIC}$ and $\rm[Fe/H]_{T23}$, as shown in the top panel of Figure \ref{fig:adj_proc}. Bins with less than ten stars receive the offset calculated in the closest bin. Then, these values were used to correct $\rm[Fe/H]_{MAGIC}$, resulting in the residual distribution presented in the middle panel of Figure \ref{fig:adj_proc}.
The bottom panel of Figure \ref{fig:adj_proc} presents the MDF obtained from the MAGIC metallicities after adjusting to align with \citetalias{Tolstoy2023} metallicity estimates. 
We emphasize that this procedure is intended to bring MAGIC metallicities to the same scale as \citetalias{Tolstoy2023}, allowing us to isolate the impact of spatial bias, however it does not represent our final adopted metallicity estimates. After this adjustment, we still find a more metal-rich peak, with a median value of $\rm[Fe/H]_{adjusted} = -1.61$ and a median absolute deviation of 0.37 dex, which indicates that the differences in the MDFs and metallicity gradients of each work is due to the spatial coverage, rather than to systematics between metallicity scales.

Our method performs well on the lower metallicity regime, potentially estimating metallicities below $\rm[Fe/H] = -3.5$. We find 135 members with $\rm[Fe/H]_{MAGIC} < -2.5$, nearly doubling the number of stars in this regime when compared to \citetalias{Tolstoy2023} (74 stars). 
Furthermore, going to even lower metallicities, we can identify 20 EMP stars. Among the stars in common with MAGIC observations, \citetalias{Tolstoy2023} found 12 EMP stars, and seven of them are also identified as EMP by MAGIC metallicity. 
Five other stars would be classified as possible EMP stars by MAGIC metallicities, however they appear to be more metal-rich by \citetalias{Tolstoy2023}, with $-2.7 \leq \rm [Fe/H]_{T23} \leq -2.4$. This represents a success rate of $\sim 58\%$ for the detection of EMP stars in our sample. Notably, we also find six stars in the Scl dSph with $\rm [Fe/H]_{MAGIC} < -3$ with no counterpart in the literature. However, these stars are faint ($G > 19$), making them feasible targets for the next generation of spectroscopic surveys.

\begin{figure}[t]
\begin{center}
\includegraphics[width=\columnwidth]{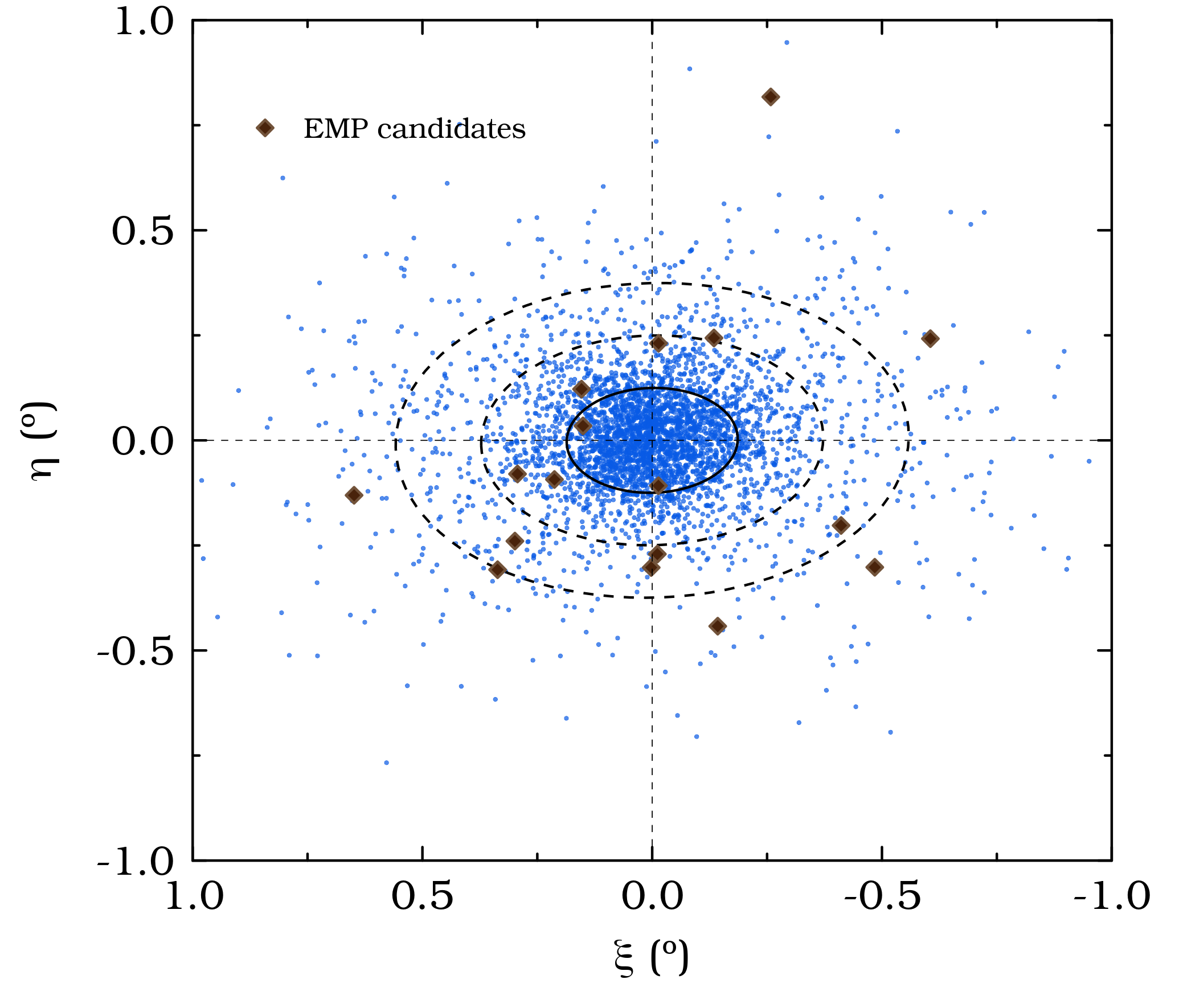}
\caption{Spatial distribution of Scl dSph members (small dots) and EMP candidates ($\rm[Fe/H]_{MAGIC} < -3.0$; diamonds) from our MAGIC photometric metallicity data.}
\label{fig:scl_emp}
\end{center}
\end{figure}

Recently, \cite{Arroyo2024} suggested the existence of a third chemodynamical stellar population in the Scl dSph galaxy, characterized by a higher mean line-of-sight velocity, a very low metallicity ($\rm [Fe/H] < -2$), and a distinctly asymmetric spatial distribution. Here, we recover this peculiar behaviour in the sky projection of the MAGIC EMP candidates. In Figure \ref{fig:scl_emp}, we highlight 17 EMP stars identified and note that most of them (11) are preferentially found on the southern region of the galaxy. Three EMP candidates were removed as they possess velocities measured by \citetalias{Tolstoy2023} below $110 \ \rm km \, s^{-1}$ and certainly do not compose this group according to \cite{Arroyo2024}'s classification. Further radial velocity analysis is required to confirm whether these candidates, as well as new very metal-poor stars observed by MAGIC, are members of this newly discovered Scl dSph population.

\begin{table}[ht]
\caption{Summary of statistics for the MAGIC MDF (Figure \ref{fig:mdf}), including mean, median, standard deviation (SD), median absolute deviation (MAD), skewness, kurtosis, and quantiles (25th and 75th). Mean and median values with their respective deviation for the normalized difference (ND) histograms (Figure \ref{fig:norm_hist}) are listed here. The metallicity gradients estimated by single and segmented fits (SF) are also presented.} \label{main_tab}
\centering
\begin{tabular}{lc}
\hline
  \multicolumn{2}{c}{MAGIC MDF} \\
\hline \hline
  Median $\pm$ MAD & $-1.65 \pm 0.42$ \\ 
  Mean $\pm$ SD & $-1.67 \pm 0.46$ \\
  Skewness & 0.03 \\
  Kurtosis & 3.98 \\
  25th / 75th quantiles & $-1.97$ / $-1.39$ \\
  \hline
  \multicolumn{2}{c}{ND: Tolstoy et al. -- $\frac{\rm [Fe/H]_{MAGIC} - [Fe/H]_{T23} - C}{\sqrt{\sigma^2_{\rm T23} + \sigma_{\rm MAGIC}^2}}$} \\[0.2cm]
  \hline \hline
  Median $\pm$ MAD & $0.0 \pm 1.1$ \\
  Mean $\pm$ SD & $0.0 \pm 1.3$ \\
  \hline
  \multicolumn{2}{c}{ND: de los Reyes et al. -- $\frac{\rm [Fe/H]_{MAGIC} - [Fe/H]_{R22} - C}{\sqrt{\sigma^2_{\rm R22} + \sigma_{\rm MAGIC}^2}} $} \\[0.2cm]
  \hline \hline
  Median $\pm$ MAD & $0.0 \pm 1.1$ \\
  Mean $\pm$ SD & $0.1 \pm 1.2$ \\
  \hline
  \multicolumn{2}{c}{Metallicity gradients } \\
  \hline \hline
  \multirow{3}{4cm}{Single linear fit} & $-1.00 \pm 0.03 \rm \ dex \, deg^{-1}$ \\
    & $-0.19 \pm 0.01 \rm \ dex \, r_h^{-1}$ \\
    & $-0.68 \pm 0.18 \rm \ dex \, kpc^{-1}$ \\[0.15cm]
  \multirow{3}{4cm}{SF: until $1.06 \ \rm r_h$} & $-3.26 \pm 0.18 \rm \ dex \, deg^{-1}$ \\
   & $-0.61 \pm 0.03 \rm \ dex \, r_h^{-1}$ \\
    & $-2.22 \pm 0.60 \rm \ dex \, kpc^{-1}$ \\[0.15cm]
  \multirow{3}{4cm}{SF: beyond $1.06 \ \rm r_h$} & $-0.55 \pm 0.26 \rm \ dex \, deg^{-1}$ \\
   & $-0.10 \pm 0.05 \rm \ dex \, r_h^{-1}$ \\
    & $-0.38 \pm 0.20 \rm \ dex \, kpc^{-1}$ \\
\hline \hline 
\end{tabular}
\end{table}

\subsection{Metallicity gradient}
\label{grad}

Numerous works have demonstrated that dwarf galaxies usually cannot be modeled by a single stellar component (e.g., \citealt{Ural2010_mp}, \citealt{deBoer2012_mp}, \citealt{Pace2020_mp}), and the Scl dSph is not an exception. Observing horizontal-branch stars, \cite{Tolstoy2004} revealed that most of the redder (more metal-rich and younger) stars tend to be found in the inner region, whereas the blue ones (more metal-poor and older) are found in the outskirts. Since then, the metallicity gradient of the Scl dSph has been quantified with more robust samples (\citealt{Kirby2009, Kirby2011}, \citealt{Leaman2013_grad}, \citealt{Ho2015_grad}, \citealt{Martinez2016_grad}, \citealt{Taibi2022}, \citealt{Tolstoy2023}). 

\begin{figure}[t]
\centering
  \includegraphics[width=\columnwidth]{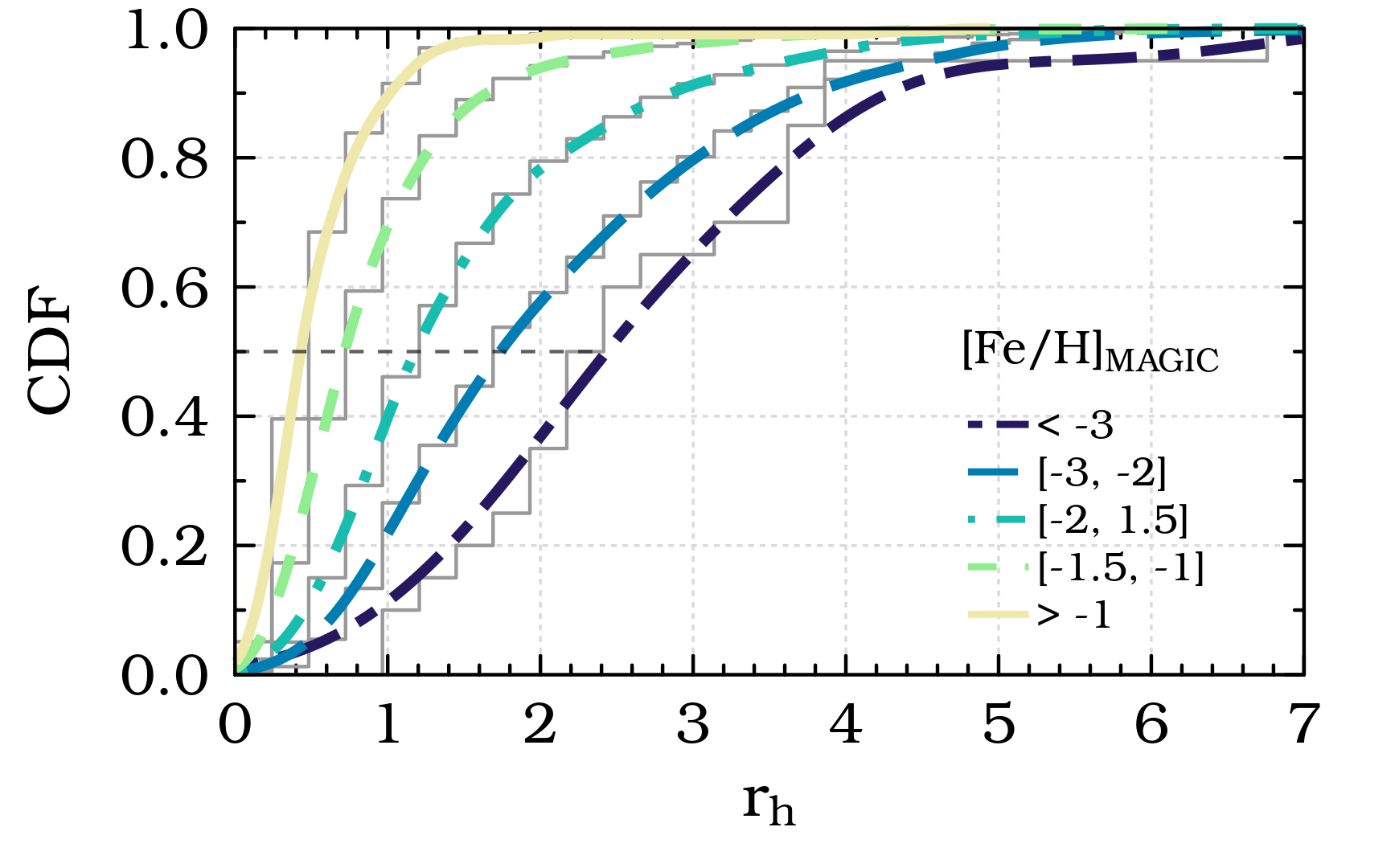}
  \caption{Cumulative distribution functions for the Scl dSph members at different metallicity intervals.}
  \label{fig:cdf_feh}
\end{figure}

Here, we quantify the metallicity gradient of the Scl dSph with MAGIC data. Figure \ref{fig:cdf_feh} presents the cumulative distribution function in units of $\rm r_h$, which reflects the metalicity gradient of the galaxy. We adopted $\rm r_h = 11.17'$, as provided by \cite{Munoz2018}. The metallicity intervals shown contain 20, 878, 1588, 1162, 235 stars from $\rm [Fe/H]_{MAGIC} < -3$ to $\rm [Fe/H]_{MAGIC} > -1$, respectively. The distributions show that more metal-rich stars are found mainly in the inner regions, while more metal-poor stars preferentially populate the outermost part of the galaxy. Beyond $\sim 3 \ \rm r_h$, we find few stars with $\rm[Fe/H]_{MAGIC} > -1.5$. 
The dashed gray line indicates the points in each subsample where we find 50\% of the stars, corresponding to the $2.4$, $1.8$, $1.2$, $0.7$, and $0.4 \rm \ r_h$. The curves were smoothed using a Gaussian kernel with a bandwidth following Silverman's rule of thumb \citep{silverman1986density}.

\begin{figure*}[pt!]
\begin{center}
\includegraphics[width=1.8\columnwidth, trim={0 1.8cm 0 0}, clip]{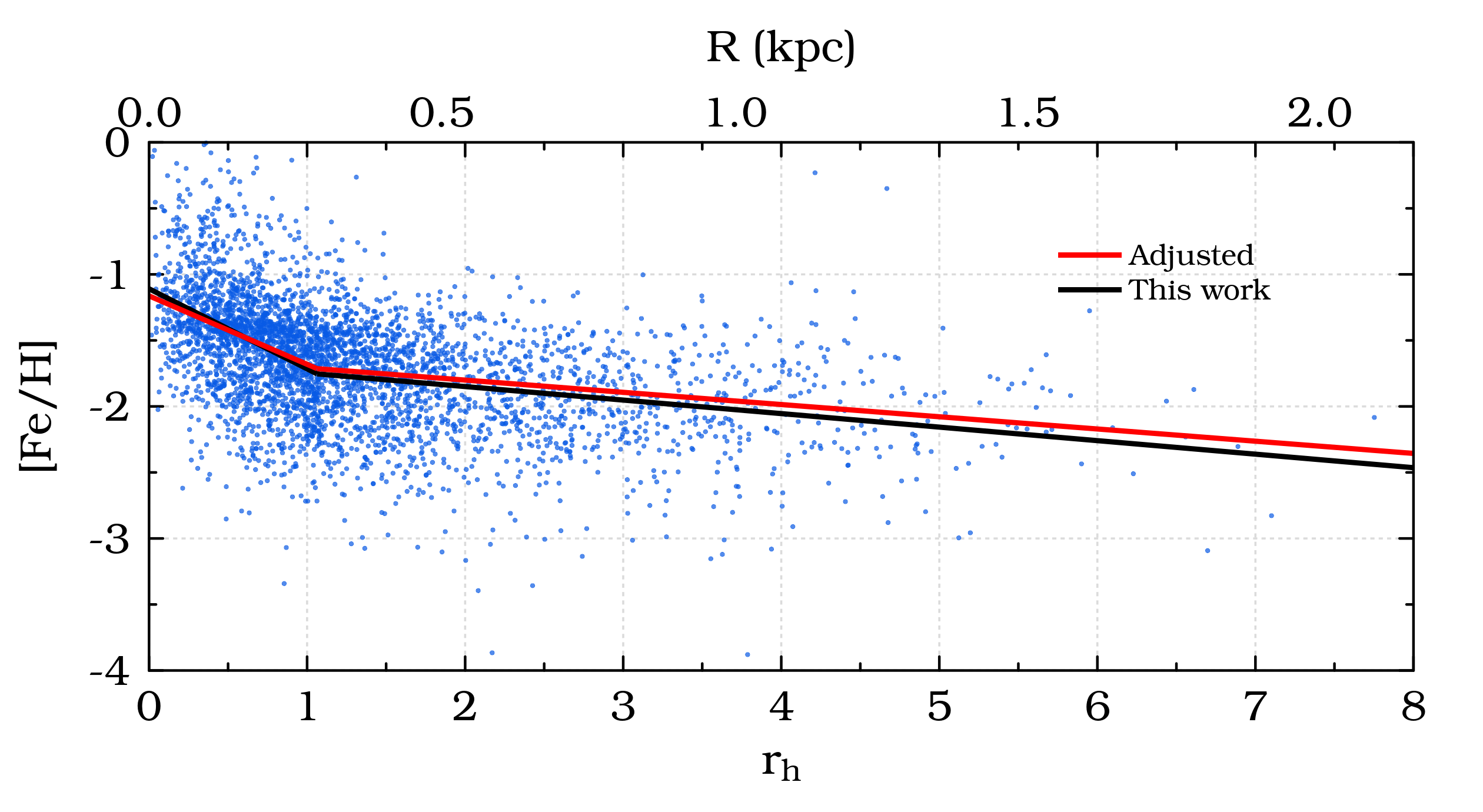}
\includegraphics[width=1.8\columnwidth]{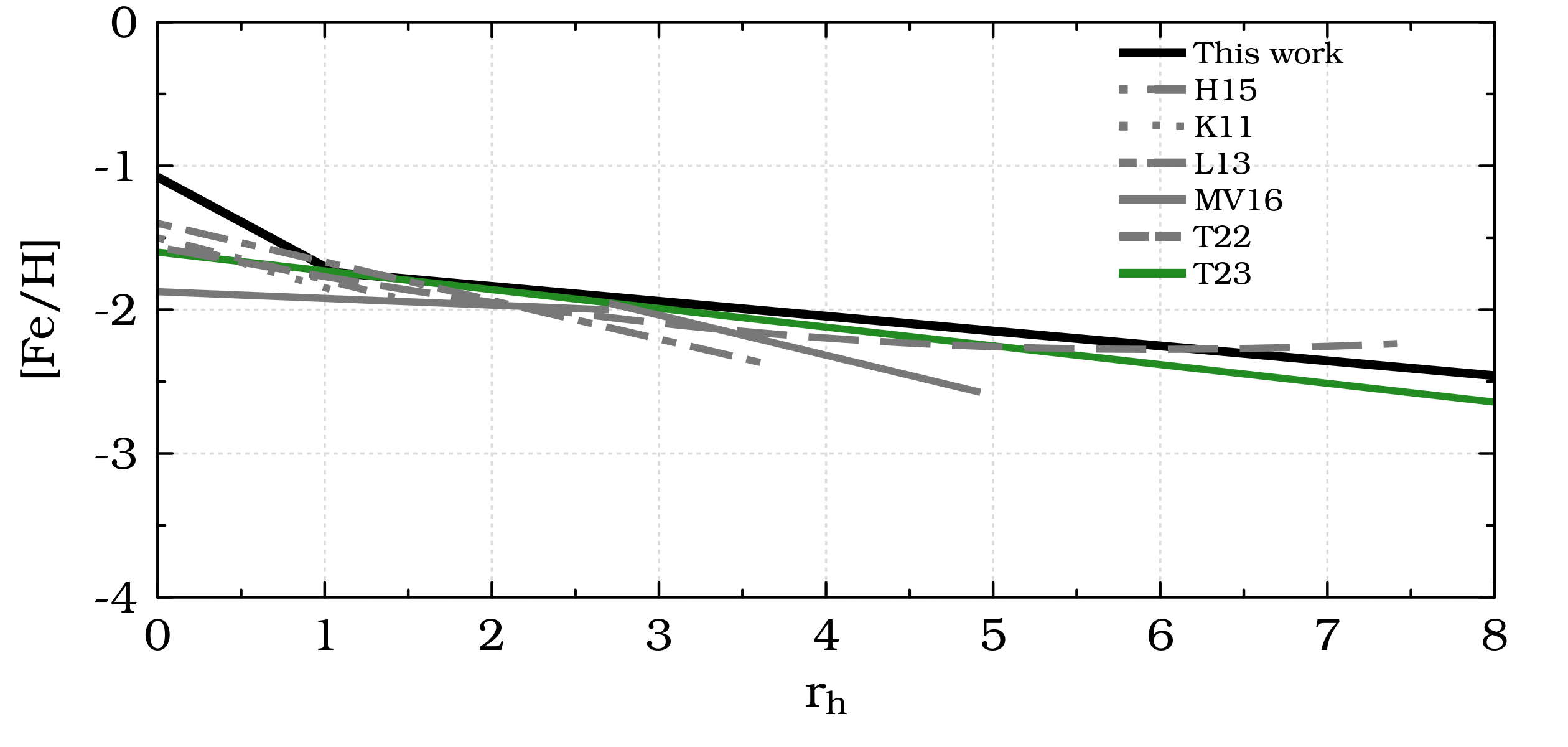}
\caption{Top panel: metallicities of the Scl dSph members as a function of half-light radius/projected distance. Black points represent median $\rm[Fe/H]_{MAGIC}$ values in bins of $\rm 0.5 \ r_h$. The black line represents a piecewise fit (inner fit: slope $= -0.62$ and intercept $=-1.08$; outer fit: slope $=-0.10$ and intercept $=-1.63$). The red line shows the segmented fit performed for adjusted MAGIC metallicities, brought to \citetalias{Tolstoy2023} scale. Bottom panel: metallicity gradients obtained in this work, along with previous estimates reported in the literature. We present metallicity gradients from \citeauthor{Kirby2011} (\citeyear{Kirby2011}; K11), \citeauthor{Leaman2013_grad} (\citeyear{Leaman2013_grad}; L13), \citeauthor{Ho2015_grad} (\citeyear{Ho2015_grad}; H15), \citeauthor{Martinez2016_grad} (\citeyear{Martinez2016_grad}; MV16), \citeauthor{Taibi2022} (\citeyear{Taibi2022}; T22), and \citeauthor{Tolstoy2023} (\citeyear{Tolstoy2023}; T23).}
\label{fig:grad}
\end{center}
\end{figure*}

Figure \ref{fig:grad} shows the variation of metallicity with $\rm r_h$ with MAGIC data (top panel) and a comparison of the metallicity gradients reported in the literature (bottom panel). 
The black line illustrates a piecewise linear fit to the MAGIC metallicities obtained using the \texttt{segmented} R package \citep{Muggeo2003_piecewise}. It evidences a sharp decrease in metallicity until $1.06 \pm 0.04 \ \rm r_h$ ($\sim 11.8'$), followed by an almost constant trend. 
The first slope measures a gradient of $-3.26 \pm 0.18 \rm \ dex \, deg^{-1}$ and, after the break point, we find a shallower slope of $-0.55 \pm 0.26 \rm \ dex \, deg^{-1}$. 
The red line indicates the segmented fit obtained for MAGIC metallicities adjusted to the \citetalias{Tolstoy2023} metallicity scale by the median offsets in Figure \ref{fig:adj_proc} (see discussion in Section \ref{mdf}). It slightly changes to $-2.79 \pm 0.16 \rm \ dex \, deg^{-1}$ and $-0.50 \ \pm \ 0.23 \rm \ dex \, deg^{-1}$, with a break-point at $1.06 \ \pm \ 0.04 \ \rm r_h$.

A double fit, with a stronger inner gradient, is also detectable in spectroscopic data. When applying the segmented fit to \citetalias{Tolstoy2023} data, we find an inner gradient of $-1.51 \pm 0.24 \rm \ dex \, deg^{-1}$ and a flatter gradient beyond $1.35 \ \pm \ 0.19 \ \rm r_h$, with a value of $-0.38 \ \pm \ 0.36 \rm \ dex \, deg^{-1}$. This result reiterates that the inner region of the Scl dSph is more chemically enriched compared to the outer region (e.g., \citealt{Martinez2016_grad}, \citealt{Bettinelli2019}). The outer gradients obtained with both MAGIC and \citetalias{Tolstoy2023} are both weaker compared with previous estimates in the literature ($-1.44 \rm \ dex \ deg^{-1}$; \citealt{Leaman2013_grad}, and $-1.50 \rm \ dex \ deg^{-1}$ for RR Lyrae stars beyond $32'$; \citealt{Martinez2016_grad}).

A single linear fit to MAGIC metallicities results in a gradient of $-1.00 \pm 0.03  \rm \ dex \, deg^{-1}$ ($-0.68 \pm 0.18 \rm \ dex \, kpc^{-1}$). The trend is expected to be stronger with our estimates compared to \citetalias{Tolstoy2023}, as our sample comprises more stars in the center of the galaxy. 
The metallicity gradient presented by \citetalias{Tolstoy2023}, of $-0.70 \rm \ dex \ deg^{-1}$, agrees very well with the estimation from \cite{Taibi2022} ($-0.68 \pm 0.05 \rm \ dex \, deg^{-1}$). Our gradient highly exceeds those measurements and might be inflated by the metallicity overestimation observed for $\rm[Fe/H]_{MAGIC} > -1$. Nevertheless, when only stars without photometric de-blending are considered, our data still implies a slightly stronger metallicity gradient of $-0.77 \pm 0.04 \rm \ dex \, deg^{-1}$. This value is closer to those presented in the literature, showing the effect of spatial biasing when excluding sources in the center that were de-blended.
If the MAGIC metallicities are adjusted to the \citetalias{Tolstoy2023} metallicity scale by the median offsets in bins of 0.2 dex, the linear gradient decreases to $-0.88 \pm 0.03 \rm \ dex \, deg^{-1}$ ($-0.60 \pm 0.16 \rm \ dex \, kpc^{-1}$). This value also indicates a stronger metallicity gradient than the fit obtained with \citetalias{Tolstoy2023} data, reinforcing that this is a result of our sample being spatially unbiased.

The origin of metallicity gradients is still unclear. It could be a consequence of secular evolution, with more metal-rich stars being formed at the central region, where the potential is stronger and gas density is higher (\citealt{Kennicutt1983_grad}, \citealt{Schroyen2013_grad}, \citealt{Revaz2018}).
Another possibility is that, with successive rounds of star formation over time, earlier generations of stars are pushed to the outskirts by supernova feedback (\citealt{El-Badry2016}, \citealt{Mercado2021}). 
Moreover, the formation of gradients is influenced by both internal and environmental phenomena. Accretion events may either strengthen or diminish it (e.g., \citealt{Benitez2016_grad}, \citealt{Cardona-barrero2021}, \citealt{Taibi2022}). By removing the gas in the outer parts, tidal stripping creates a less enriched environment \citep{WMRomeo2016}. Additionally, the rotation of the galaxy can prevent the gas from sinking to the center, resulting in shallower metallicity gradients (\citealt{Schroyen2011}, \citealt{Leaman2013_grad}).
Just recently, observations of metallicity variations along the stream of the Sagittarius dSph (e.g., \citealt{Majewski2003}) have been linked to a rapid formation of a metallicity gradient in this galaxy, which should have been in place already at redshift z $\sim 0.5$ (\citealt{Limberg2023}, \citealt{Cunningham2024}).

Strong metallicity gradients were previously reported for the Scl dSph (e.g., $-1.86 \rm \ dex \ deg^{-1}$ by \citealt{Kirby2011} and $-1.57 \rm \ dex \ deg^{-1}$ by \citealt{Ho2015_grad}) and are indeed expected for systems with distinct chemo-dynamical populations. 
\cite{Benitez2016_grad} discussed the possible role of mergers in the formation of the Scl dSph, being responsible for the spatial segregation of the metal-rich and metal-poor populations. Minor mergers might contribute to the less pronounced gradients in the outer region of the galaxy. Theoretical works analyzing cosmological simulations indicate that mergers in dwarf galaxies are not unusual (\citealt{Deason2014}, \citealt{Deason2022}). Dwarf galaxies are expected to experience $\sim1$ major merger, and between 5--9 minor mergers (\citealt{Fitts2018}, \citealt{Deason2022}).
\cite{Arroyo2024} argue that the asymmetric spatial distribution of EMP stars could be the result of a recent merger event in the Scl dSph, supported by the shift in line-of-sigth velocity and the dispersion of metallicity. In this study, the smooth decline beyond 1.06 $\rm r_h$ suggests a weak gradient in the outskirts, however, it is not definitive evidence for a merger. To verify this hypothesis, more research should be done on the outskirts of the Scl dSph using high-resolution spectroscopy and detailed abundance analysis.

\section{Conclusions \& Summary}
\label{sec:conc}

This work presents a comprehensive study of the photometric metallicities of stars in the Scl dSph galaxy from the DECam MAGIC survey, including 3883 member stars out to seven times its half-light radius. Our study roughly triples the numbers of stars in the literature with spectroscopic metallicities, and allows us to derive the metallicity properties of the system in a spatially unbiased manner into the far outskirts of the system. Given the Scl dSph’s status as a ``textbook'' dwarf galaxy that is well-studied, we use the extensive literature data to validate our photometric metallicity estimates and to assess the utility of our program on other dwarf galaxies in the near future.

We find a median metallicity of -1.65 dex for the Scl dSph, and the metallicity estimates derived from the CaHK filter agree well with spectroscopic data, especially between $-2.5 < \rm[Fe/H]_{MAGIC} < -1.0$, with a median offset around $-0.05$ dex. For stars more metal-rich than $-1$ dex, MAGIC metallicities tend to be higher, with an offset of $\sim 0.32$ dex, while the opposite is observed for stars with $\rm[Fe/H]_{MAGIC} < -2.5$ (offset $\sim -0.18$ dex). When comparing only stars in common, our estimates recover an MDF very similar to the one obtained by Ca II triplet metallicities in the \citetalias{Tolstoy2023} study of $\sim 1400$ stars, with a median value of $\rm[Fe/H]_{MAGIC} = -1.84$ and a median absolute deviation of $0.44$ dex. The agreement with spectroscopic values demonstrates the reliability of our measurements. 

Our study shows that the MDF of the Scl dSph galaxy exhibits a characteristic expected for dSph systems: a sharp drop at the metal-rich end, an indication of star formation interrupted in dSph satellites by ram pressure stripping (\citealt{Kirby2013}, \citealt{Ross2015}). 
The radial distribution shows a clear change in the spatial distribution of stars according to their metallicities, with the radii enclosing approximately 50\% of the stars increasing from $0.4 \rm \ r_h$ to $2.4 \rm \ r_h$ for those with $\rm[Fe/H]_{MAGIC} > -1$ and $\rm[Fe/H]_{MAGIC} < -3$, respectively. 
We estimated a metallicity gradient of $-3.26 \pm 0.18 \rm \ dex \, deg^{-1}$ for stars inside $1.06 \ \rm r_h$ and $-0.55 \pm 0.26 \rm \ dex \, deg^{-1}$ for the outer part of the galaxy. The internal metallicity gradient is stronger than previously estimated, incompatible with values reported by recent spectroscopic works (\citealt{Taibi2022} and \citealt{Tolstoy2023}), which is likely due to the lack of more metal-rich stars found in the center of the Scl dSph in previous samples. However, it is important to clarify that this value is affected by the apparently worse performance noticed for $\rm[Fe/H]_{MAGIC} > -1$ in the MAGIC photometric metallicity estimates, even though we still expect a stronger gradient when stars in central regions are accessed.

The MAGIC photometric metallicities exhibit promising performance, even into the lowest metallicity regimes, as observed in the bottom panel of Figure \ref{fig:mdf}. This allows us to further investigate the low metallicity population in our sample. Comparing with \citetalias{Tolstoy2023} spectroscopic data, we obtain a success rate of $\sim 58\%$ in identifying stars below $\rm [Fe/H] = -3$. Furthermore, we found six EMP candidates without spectroscopic confirmation in the literature. All the stars present magnitudes $G > 19$, which makes it difficult to perform spectroscopic follow-up using facilities currently in operation. Still, it is another example of how investigating photometric metallicities can be useful in distant galaxies.  
Additional studies in the outer region in the Scl dSph and other dwarf galaxies are essential to find the most metal-poor stars in these systems.

This paper provides a demonstration of the power of the photometric metallicity technique with the CaHK narrow-band filter used by the MAGIC survey, building on the success of photometric metallicity studies of dwarf galaxies in the past (e.g., \citealt{Chiti2020, Chiti2021}, \citealt{Vitali2022}, \citealt{Longeard2022}, \citealt{Fu2023, Fu2024_M31}, \citealt{Pan2025}). Notably, we can obtain much larger samples of members for dwarf galaxies without target-selection biases introduced by spectroscopic studies. In the future, we will expand this analysis to other Local Group dSph galaxies less well-studied than the Scl dSph, providing important population-level constraints regarding their evolution through the properties of their stellar populations (e.g., MDFs, spatial-metallicity correlations). 
The MAGIC survey is still ongoing, with an upcoming early science paper (Chiti et al. in prep.), an initial exploration of the detailed chemistry of low metallicity stars in the outer halo \citep{Placco2025}, and upcoming work relevant to dwarf galaxies and associated low-metallicity substructures and streams.
Our initial study serves as a powerful demonstration and validation of the utility of MAGIC photometric metallicities, to significantly enhance such studies of resolved stellar populations accross the Milky Way ecosystem.

\acknowledgments

We thank the referee for the review and suggestions that helped to improve our manuscript. 
Thanks to S. Taibi, who kindly provided the metallicity gradient obtained with the GPR method, and R. M. Santucci for the comments during the development of this work. 
This study was financed, in part, by the São Paulo Research Foundation (FAPESP), Brasil. Process Number \#2022/16502-4 and \#2020/15245-2. 
G.L. acknowledges support from KICP/UChicago through a KICP Postdoctoral Fellowship. 
W.C. gratefully acknowledges support from a Gruber Science Fellowship at Yale University. This material is based upon work supported by the National Science Foundation Graduate Research Fellowship Program under Grant No. DGE2139841. Any opinions, findings, and conclusions or recommendations expressed in this material are those of the author(s) and do not necessarily reflect the views of the National Science Foundation.  
The work of V.M.P is supported by NOIRLab, which is managed by the Association of Universities for Research in Astronomy (AURA) under a cooperative agreement with the U.S. National Science Foundation. 
A.P.J. acknowledges support from NSF grant AST-2307599. 
JAC-B acknowledges support from FONDECYT Regular N 1220083.

The DELVE project is partially supported by the NASA Fermi Guest Investigator Program Cycle 9 No. 91201.
This work is partially supported by Fermilab LDRD project L2019-011. 
This material is based upon work supported by the National Science Foundation under Grant No. AST-2108168, AST-2108169, AST-2307126, and AST-2407526.

This project used data obtained with the Dark Energy Camera (DECam), which was constructed by the Dark Energy Survey (DES) collaboration. 
Funding for the DES Projects has been provided by the US Department of Energy, the U.S. National Science Foundation, the Ministry of Science and Education of Spain, the Science and Technology Facilities Council of the United Kingdom, the Higher Education Funding Council for England, the National Center for Supercomputing Applications at the University of Illinois at Urbana–Champaign, the Kavli Institute for Cosmological Physics at the University of Chicago, the Center for Cosmology and Astro-Particle Physics at the Ohio State University, the Mitchell Institute for Fundamental Physics and Astronomy at Texas A\&M University, Financiadora de Estudos e Projetos, Fundação Carlos Chagas Filho de Amparo à Pesquisa do Estado do Rio de Janeiro, Conselho 12 Nacional de Desenvolvimento Científico e Tecnológico and the Ministério da Ciência, Tecnologia e Inovação, the Deutsche Forschungsgemeinschaft and the Collaborating Institutions in the Dark Energy Survey.

The Collaborating Institutions are Argonne National Laboratory, the University of California at Santa Cruz, the University of Cambridge, Centro de Investigaciones Enérgeticas, Medioambientales y Tecnológicas–Madrid, the University of Chicago, University College London, the DES-Brazil Consortium, the University of Edinburgh, the Eidgenössische Technische Hochschule (ETH) Zürich, Fermi National Accelerator Laboratory, the University of Illinois at Urbana-Champaign, the Institut de Ciències de l’Espai (IEEC/CSIC), the Institut de Física d’Altes Energies, Lawrence Berkeley National Laboratory, the Ludwig-Maximilians Universität München and the associated Excellence Cluster Universe, the University of Michigan, NSF NOIRLab, the University of Nottingham, the Ohio State University, the OzDES Membership Consortium, the University of Pennsylvania, the University of Portsmouth, SLAC National Accelerator Laboratory, Stanford University, the University of Sussex, and Texas A\&M University.

Based on observations at NSF Cerro Tololo Inter-American Observatory, NSF NOIRLab (NOIRLab Prop. ID 2019A-0305; PI: Alex Drlica-Wagner, and NOIRLab Prop. ID 2023B-646244; PI: Anirudh Chiti), which is managed by the Association of Universities for Research in Astronomy (AURA) under a cooperative agreement with the U.S. National Science Foundation.

This manuscript has been authored by Fermi Research Alliance, LLC under Contract No. DE-AC02-07CH11359 with the U.S. Department of Energy, Office of Science, Office of High Energy Physics. The United States Government retains and the publisher, by accepting the article for publication, acknowledges that the United States Government retains a non-exclusive, paid-up, irrevocable, world-wide license to publish or reproduce the published form of this manuscript, or allow others to do so, for United States Government purposes.

This work has made use of data from the European Space Agency (ESA) mission {\it Gaia} (\url{https://www.cosmos.esa.int/gaia}), processed by the {\it Gaia} Data Processing and Analysis Consortium (DPAC, \url{https://www.cosmos.esa.int/web/gaia/dpac/consortium}). Funding for the DPAC has been provided by national institutions, in particular the institutions participating in the {\it Gaia} Multilateral Agreement.

This research made use of RStudio \citep{rstudio}, TOPCAT (\url{http://www.starlink.ac.uk/topcat/}, \citealp{topcat}), and VizieR catalogue access tool, CDS, Strasbourg, France \citep{10.26093/cds/vizier}. The original description of the VizieR service was published in \citet{vizier2000}. This work also made use of NASA's Astrophysics Data System Bibliographic Services and of the VALD database, operated at Uppsala University, the Institute of Astronomy RAS in Moscow, and the University of Vienna.

\bibliography{bibliography}{}
\bibliographystyle{aasjournal}

\facilities{Blanco (DECam), Astro Data Lab, Gaia}

\software{RStudio \citep{rstudio},  
          Scipy \citep{2020SciPy},
          TOPCAT \citep{topcat}
          }

\appendix

\section{Comparison with spectroscopic works}
\label{appen}

In Table \ref{tab:stat}, we compile the median values and offsets obtained comparing MAGIC estimates to previous datasets. The \cite{Chiti2018} study gives the worst comparison to the photometric metallicities, since their sample generally includes lower metallicity stars that are more carbon-enhanced. As mentioned in Section \ref{sec:comp}, the accuracy of photometric [Fe/H] decreases to more metal-poor stars and carbon enhancement can also affect the estimates. In addiction to [Fe/H], we included [Ca/H] values from \citetalias{delosReyes22}, since the MAGIC metallicites are derived from flux measured in the Ca H and K lines. Although the median values agree better with a similar offset, the distribution observed for [Ca/H] is not consistent with the MDF presented in Section \ref{mdf}.  
\begin{table}[ht]
\caption{Comparison of MAGIC data with spectroscopic samples. For each work, the number of stars in common (N) and the median values using the spectroscopic ($\rm[Fe/H]_{spec}$) and photometric ($\rm[Fe/H]_{MAGIC}$) estimates are presented. The median values of $\Delta\rm[Fe/H]$ are also provided ($\rm M_{\Delta[Fe/H]}$).} \label{tab:stat}
\centering
\begin{tabular}{lcccc}
\hline
  \multicolumn{1}{c}{Work} &
  \multicolumn{1}{c}{N} &
  \multicolumn{1}{c}{$\rm[Fe/H]_{spec}$} &
  \multicolumn{1}{c}{$\rm[Fe/H]_{MAGIC}$} & 
  \multicolumn{1}{c}{$\rm M_{\Delta[Fe/H]}$} \\
\hline \hline
  Tolstoy et al. (2023) & 1148 & -1.82 & -1.84 & -0.05 \\
  Walker et al. (2023) & 194 & -1.59 & -1.77 & -0.21 \\
  de los Reyes et al. (2022) -- [Fe/H] & 366 & -1.66 & -1.57 & 0.04 \\
  de los Reyes et al. (2022) -- [Ca/H] & 350 & -1.50 & -1.57 & -0.16 \\
  Reichert et al. (2020) & 77 & -1.56 & -1.79 & -0.21 \\
  Hill et al. (2019) & 76 & -1.57 & -1.74 & -0.17 \\
  Chiti et al. (2018) & 64 & -2.84 & -2.35 & 0.38 \\
  Kirby et al. (2009) & 328 & -1.58 & -1.55 & 0.04 \\
  APOGEE DR17 \citep{Abdurrouf2022_apogee} & 147 & -1.69 & -1.97 & -0.24 \\
\hline \hline 
\end{tabular}
\end{table}

\section{Data availability}
\label{appen2}

The catalog for Scl dSph members analyzed in this work is available at the following Zenodo DOI: \href{https://zenodo.org/records/16326635}{10.5281/zenodo.16326635} \citep{barbosa_2025_16326635}. It presents both Gaia and DELVE photometry, membership probabilities from \cite{Pace2022}, and stellar parameters ($\log \rm g$ and [Fe/H]) derived following the method descripted in Section \ref{feh}. A fragment of the full online sample is presented in Table \ref{tab:cat}.

\begin{table}[ht]
\centering
\caption{Scl dSph member analyzed in this work. Gaia source\_id, right ascension (ra), declination (dec), and \textit{G} magnitude are provided. Photometry from DELVE DR2 (\textit{g}-, \textit{r}-, and \textit{i}-band) are available, as well as derived $\log \ \rm g$ and [Fe/H].} 
\begin{tabular}{ccccccccc}
\hline 
  \multicolumn{1}{c}{\texttt{source\_id}} &
  \multicolumn{1}{c}{\texttt{ra}} &
  \multicolumn{1}{c}{\texttt{dec}} &
  \multicolumn{1}{c}{\texttt{phot\_g\_mean\_mag}} &
  \multicolumn{1}{c}{\texttt{g\_dered}} &
  \multicolumn{1}{c}{\texttt{r\_dered}} &
  \multicolumn{1}{c}{\texttt{i\_dered}} &
  \multicolumn{1}{c}{\texttt{logg\_magic}} &
  \multicolumn{1}{c}{\texttt{feh\_magic}} \\
\hline \hline
  5003194866102866048 & 15.006 & -33.855 & 17.470 & 18.157 & 17.368 & 17.064 & 1.155 & -2.221\\
  5003199775249111680 & 14.955 & -33.735 & 17.762 & 18.420 & 17.628 & 17.324 & 1.203 & -1.709\\
  5027218267457408384 & 15.031 & -33.722 & 17.249 & 17.989 & 17.102 & 16.775 & 0.936 & -1.891\\
  5027219126450866816 & 15.010 & -33.688 & 18.059 & 18.729 & 17.917 & 17.637 & 1.357 & -1.126\\
  5027219779285852672 & 15.010 & -33.655 & 18.103 & 18.708 & 17.974 & 17.709 & 1.420 & -1.595\\
\hline \hline
\end{tabular}
\label{tab:cat}
\end{table}

\end{document}